\documentclass[12pt]{iopart}

\makeatletter
\let\@original@xhline\@xhline
\usepackage{array}
\let\@xhline\@original@xhline
\makeatother

\expandafter\let\csname equation*\endcsname\relax
\expandafter\let\csname endequation*\endcsname\relax
\usepackage{amsmath,amssymb,amsthm,enumerate,graphicx,pgf,tikz,fancyhdr}
\usepackage[utf8]{inputenc}
\usetikzlibrary{arrows}

\usepackage{lipsum}
\usepackage{tabularx}

\usepackage{dcolumn}
\usepackage{bm}

\usepackage{hyperref}
\usepackage[numbers,square,sort&compress]{natbib}
\newtheorem{theorem}{Theorem}
\newtheorem{Definition}{Definition}

\newtheorem{Proposition}{Proposition}
\newtheorem{lemma}{Lemma}

\newtheorem{Protocol}{Protocol}

\usepackage{ragged2e,array,booktabs}

\usepackage{tcolorbox}
\usepackage{bbm}

\usepackage{comment}

\usepackage{dashbox}
\usepackage{todonotes}
\usetikzlibrary{shapes.callouts}
\usepackage{csquotes}
\usepackage{multirow}

\usepackage[toc,page]{appendix}

\definecolor{ss_color}{rgb}{1,0,0}

\definecolor{blue}{rgb}{0,0,1}

\hypersetup{hidelinks,pdftitle={Everlasting secure key agreement from a quantum computational hybrid security model (V7.3)},pdfsubject={V7.3: revisions relative to V6 marked in blue}}
\AtBeginEnvironment{thebibliography}{%
  \let\VsevenOriginalBibitem\bibitem
  \renewcommand{\bibitem}[2][]{%
    \par\def\VsevenCurrentKey{#2}\def\VsevenNewKey{HayashiNagaoka2003}\def\VsevenHMKey{Mazzoncini2025hybridquantum}%
    \ifx\VsevenCurrentKey\VsevenNewKey\color{blue}\else\ifx\VsevenCurrentKey\VsevenHMKey\color{blue}\else\color{black}\fi\fi
    \VsevenOriginalBibitem[#1]{#2}}%
}

\definecolor{suggestiongreen}{rgb}{0.0,0.55,0.0}

\usepackage{bbding}
\usepackage{dsfont}

\begin{document}

\title{Everlasting secure key agreement from a quantum computational hybrid security model}

\author{Nilesh Vyas$^1$ and Romain All\'eaume$^2$}
\address{$^{1,2}$ T\'el\'ecom Paris, LTCI, Institut Polytechnique de Paris, 19 Place Marguerite Perey, 91120 Palaiseau, France}

\ead{$^1$nilesh.vyas@telecom-paris.fr, $^2$romain.alleaume@telecom-paris.fr }

\begin{abstract}
We introduce the Quantum Computational Timelock (QCT) model, combining time-lock encryption with a quantitative noisy-quantum-storage assumption. Time-lock encryption gives Bob immediate access to the encoding seed and gives Eve the same final information as access to that seed only after a protected interval. Eve may then process her retained information without computational restriction. We state this delayed-access encryption assumption explicitly and construct MUB-QCT from a complete set of mutually unbiased bases in dimension $d$ and an affine pair-wise independent permutation family.

For single-copy encoding, every seed-independent immediate measurement has trace-distance advantage at most $1/(2\sqrt{d+1})$ and mutual information at most $1/(d+1)$. Effective residual storage at diamond distance at most $\delta$ from replacement adds at most $\delta$ to the individual-attack distinguishing bound. Under i.i.d. individual attacks and entropy-preserving acceptance, we obtain an asymptotic key-rate bound and formulate final-key secrecy in composable trace distance.

For $m$ independently randomized signals sharing one bit and seed, the optimal leakage under non-adaptive product measurements is $\min\{1,m/(d+1)\}$, allowing a linear-in-$d$ detection enhancement in that attack model. We also construct a seed-independent joint measurement that recovers the bit after disclosure with error at most $\varepsilon$ from $O(\log d+\log(1/\varepsilon))$ signals, retaining only classical information after measurement. This gives a quantitative separation between product and joint measurements. Bob uses two logical detection outcomes independently of $d$.

\end{abstract}

\noindent{\it Keywords}: 
Quantum cryptography, hybrid security model, everlasting security, quantum key distribution, mutually unbiased bases

\maketitle
\section{Introduction}

\paragraph{Progress of QKD towards practical quantum cryptographic advantage:}
Quantum Key Distribution (QKD) enables key agreement with information-theoretic security, in contrast with classical key agreement protocols, whose security rests on computational hardness conjectures. QKD can therefore offer, \textit{in principle}, a distinctive security advantage over classical techniques, in particular whenever long-term protection is required. For this advantage to materialize in practice, however, two aspects must be jointly articulated: practicality (how to build efficient and cost-effective QKD systems) and security (how to guarantee an adequate security gain with respect to existing classical techniques).

Efforts are invested in making progress towards practical and secure QKD systems \cite{Advances19}. On the practicality side, QKD systems with steadily increasing performance are being developed and deployed over real-world optical networks \cite{SECOQCWP, Sasaki11, NPJ16, QKDDeployZhang18}. On the security side, a stable and robust body of work lays the foundations of the theoretical security of QKD \cite{RennerPhD, ScaraniRMP09, Toma17}, while implementation security is addressed by dedicated efforts \cite{ETSIWP, Xu19, ETSIQKDPP}, paving the way towards the certification of quantum cryptographic implementations in the near term.

\paragraph{The practical QKD dilemma:}

Despite progress over the past 25 years, a structural issue still hinders the full adoption of QKD for real-world security applications: the practicality and the security aspects of QKD are largely addressed in isolation from one another, which leads to competing requirements. On the one hand, QKD system engineering is mostly driven by performance and cost trade-offs, only part of the design constraints being derived from security considerations. On the other hand, the dominant approach to QKD security remains focused on the quest for ``ultimate security'', which, as noted by QKD pioneers \cite{GisinRMP02}, coincides with infinite cost.

The observation that contradictory goals are pursued within the QKD community was already voiced more than a decade ago by Valerio Scarani and Christian Kurtsiefer in their ``black paper on quantum cryptography'' \cite{BlackPaper}, where they stressed the need to acknowledge that practical quantum cryptography (in particular, practical QKD) and abstract quantum cryptography largely pursue different objectives.
 
This dilemma remains largely unsettled today: QKD researchers and engineers pursue a wide diversity of goals, and their views notably diverge from those of modern cryptography practitioners. This contributes to making the debate on the usefulness of QKD for real-world use cases both complex and disputed \cite{PatersonWhyQC, TCS14, McGrewPQC2015, NCSC20, AllHDR21}.

\paragraph{QCT security model as a new route towards practical quantum cryptography:}
 
To progress towards more practical quantum cryptography, we target two main objectives:

\begin{itemize}
\item Identify new constructions of quantum cryptographic primitives and protocols whose security can be formally proven, and shown to be  stronger than the security achievable by classical means alone, i.e.\ without quantum communication.

\item Focus on constructions that are implementable with existing or near-term technology, and that can therefore be demonstrated experimentally. Aim for quantum cryptographic protocols that bring a performance advantage over existing quantum cryptographic protocols proven secure in the standard model, i.e.\ without any assumption limiting the adversary.

\end{itemize} 

Our approach consists in devising a new security model, the \textit{Quantum Computational Timelock} (QCT) model, which leverages short-term computational security and noisy quantum storage in order to boost the performance and the functionality of quantum-based key agreement.

The QCT model targets \textit{everlasting secrecy}: information-theoretic secrecy against an adversary whose computational limitations are removed after the protected interval \cite{Unr13}. Our results use the time-lock encryption assumption of Definition~\ref{time-lock} together with the effective storage bound. The comparison with classical computational key agreement concerns protocols whose initial information-theoretic secret resources are smaller than the generated key.

\paragraph{The MUB-QCT construction:}

We propose the \textit{MUB-Quantum Computational Timelock} (MUB-QCT) protocol. A bit is encoded into a $d$-dimensional quantum state using MUBs and pair-wise independent permutations. For single-copy signals, we bound arbitrary seed-independent measurements and effective residual storage, and derive a conditional asymptotic rate under i.i.d. individual attacks. For multiple independently randomized signals sharing the same bit and seed, we establish the exact non-adaptive product-measurement information optimum and a logarithmic-copy joint-measurement attack. The joint attack retains only classical information until seed disclosure. Together these results quantify the security of single signals and the effect of repeating a shared bit and seed.

\section{Preliminaries}
\label{sec:preli}

We first fix the notational conventions used throughout the paper. For a positive integer $d$, we write $ [d]:=\{0,\ldots,d-1\}.$ We use uppercase italic letters to denote both classical random variables and quantum registers; the distinction will be explicitly stated or clear from the context. For classical variables, we denote probability distributions as $P_{X}$ over a finite alphabet $\mathcal{X}$, and realizations by the corresponding lowercase letter $x \in \mathcal{X}$. Vectors are denoted using superscripts: $x^{n} :=(x_{1},\ldots , x_{n})$, with $x^{n}\in \mathcal{X}^{n}$.

For a quantum system (register) $A$, the associated Hilbert space is denoted $\mathcal{H}_A$ and its dimension $|A|$; the state of a register is described by a density operator $\rho$. We use the following registers: $A$, the $d$-dimensional quantum system transmitted through the channel, which is composite, $\mathcal{H}_A = \mathcal{H}_{A_1} \otimes \mathcal{H}_{A_2}$; $A_1$, the two-dimensional subsystem carrying the encoded classical bit $x$; $A_2$, the $(d/2)$-dimensional subsystem used for randomization; and $Q$, the quantum register intercepted and stored by the adversary, Eve.

Throughout the protocol, $n$ denotes the number of rounds, $m$ the number of signals sharing one bit and seed, and $q=\log_2d$ the number of qubits needed to represent a signal. All logarithms are base two. The logarithmic copy bound below is expressed in the signal dimension: $O(\log d)=O(q)$.

For density operators $\rho$ and $\omega$, we use the trace distance $ D(\rho,\omega) := \frac{1}{2}\|\rho-\omega\|_1. $ The binary entropy is $ h_2(p):=-p\log_2p-(1-p)\log_2(1-p). $ We write $\operatorname{negl}(\lambda)$ for a function negligible in the classical security parameter $\lambda$.

\begin{Definition}[Mutually unbiased bases]
\label{MUBs}
Let $\mathcal B_{\theta_1} =\{|e_i^{\theta_1}\rangle\}_{i\in[d]},$ $\mathcal B_{\theta_2}=\{|e_j^{\theta_2}\rangle\}_{j\in[d]}$ be two orthonormal bases of a $d$-dimensional Hilbert space. They are mutually unbiased if
\begin{equation}
\left|\langle e_i^{\theta_1} |e_j^{\theta_2}\rangle\right|^2 = \frac{1}{d}
\end{equation}
for all $i,j\in[d]$.
\end{Definition}

\par In a $d$-dimensional Hilbert space, there exist at most $d+1$ mutually unbiased bases \cite{fullset}. A complete set of \(d+1\) MUBs is known to exist for prime-power dimensions. \cite{Prime}.

\begin{Definition}[Pair-wise independent permutation]
\label{permutation}
A family $\mathcal{P}$ of permutations on a set of $d$ elements, denoted $[d]$, is pair-wise independent if for all $i_{1} \neq i_{2}$ and $j_{1} \neq j_{2}$, choosing $\sigma$ uniformly at random from $\mathcal{P}$ satisfies:
\begin{equation}
\operatorname{Pr}\left\{\sigma\left(i_{1}\right)=j_{1}, \sigma\left(i_{2}\right)=j_{2}\right\}=\frac{1}{d(d-1)} .
\end{equation} 
\end{Definition}

\begin{Definition}[Time-lock encryption assumption]
\label{time-lock}
Fix a duration $t_{comp}$ and bounds on Eve's computational and parallel resources during that interval. We assume an encryption scheme $(\operatorname{Gen},\operatorname{Enc},\operatorname{Dec})$ with immediate decryption for Bob and \emph{delayed plaintext access for Eve}. Alice sends $C=\operatorname{Enc}_{k}(S)$; Bob holds an authorized key or trapdoor $k_B$ and obtains $\operatorname{Dec}_{k_B}(C)=S$ before decoding the quantum signal.

The assumed security property of the ciphertext is the following: \emph{for every admissible Eve, access to $C$ can be simulated by giving her $S$ only at time $t_{comp}$}. Before that time the simulator receives the same quantum signals as Eve and processes them using only public information and randomness independent of $S$. At $t_{comp}$ it receives $S$ and reproduces Eve's complete final information, including the retained ciphertext, decryption information, classical records and residual quantum systems. The simulation uses the same quantum-storage constraint as Eve; after $t_{comp}$ its computation is unrestricted.

We assume exact equality of the real and simulated final states, for every plaintext seed and every correlated quantum input, using one simulator for each adversary. Thus the encryption delays the usefulness of the ciphertext for Eve's quantum processing until the protected interval has elapsed. Information available directly from the quantum signals is included in both experiments.
\end{Definition}

This is the time-lock encryption property assumed in the security proofs. Section~\ref{sec:QCTmodel} expresses it mathematically. Its computational setting can be symmetric encryption with a pre-shared key, or trapdoor time-release encryption \cite{TRE,Unr06}; Appendix~\ref{CryptoDefinitions} describes both settings. The duration $t_{comp}$ is the protected interval associated with each signal.

\begin{Definition}[$(t_{comp},\delta)$-noisy quantum memory]
\label{full-decoherence}
For each admissible attack, $Q$ includes all residual quantum systems and ancillas retained by Eve. We assume that the effective input-to-disclosure storage channel $\Phi^Q_{t_{comp}}$ incorporates the allowed controls, transfers between memories, refresh operations and error correction over the protected interval. Together with time-lock encryption, the attack has the form of a seed-independent instrument with classical record $W$, followed by effective storage. We assume the bound below uniformly over admissible strategies and, for outcome-dependent storage, over the branches $W=w$.

At $t_{comp}$, the storage channel satisfies
\begin{equation}
\frac{1}{2} \left\| \Phi^Q_{t_{comp}} - \mathcal F_Q \right\|_\diamond \leq \delta,
\label{decoherence}
\end{equation}
where
\begin{equation}
\mathcal F_Q(\omega_Q) := \operatorname{Tr}[\omega_Q]\, \frac{\mathbb I_Q}{|Q|}
\end{equation}
is the completely depolarizing replacement channel on $Q$. The parameter satisfies $0\leq\delta\leq1$. A fixed replacement state other than the maximally mixed state would also suffice. For the individual-attack analysis, $\delta$ is a per-round bound on the complete residual register. A block analysis uses a storage parameter for the full joint residual register.

\end{Definition}

\noindent The coherence time $t_{coh}$ provides only a heuristic indication of long-term memory degradation and does not by itself determine the diamond-distance parameter $\delta$. Experimental storage and coherence times depend strongly on the physical platform and can range from short optical timescales to seconds or longer. The QCT security assumption is therefore stated directly in terms of the parameter $\delta$ in equation~(\ref{decoherence}), rather than through a universal relation between $t_{coh}$ and $t_{comp}$. Although decoherence may in practice be highly structured, the above definition only assumes that the best memory available to Eve is constrained by this diamond-norm distance. It also implicitly assumes that Eve's technological capabilities are limited, in the sense that she cannot implement fault-tolerant quantum error correction (QEC) with sufficient fidelity and with a low enough overhead to artificially extend the logical coherence time beyond the computational threshold $t_{comp}$.

The notion of time-limited quantum memory is closely related to the quantum storage assumed in the noisy-storage model \cite{NoisyStorage}, and time-limited quantum storage is also assumed in quantum data locking (QDL) constructions \cite{QDLDV, QDLCV}.

\section{Quantum Computational Timelock Security Model }
\label{sec:QCTmodel}
To formalize our hybrid security model, we first outline the baseline communication resources and the adversarial capabilities.

Within the QCT framework, the authorized parties, Alice and Bob, communicate via an authenticated, noiseless classical channel and an insecure quantum channel. The adversary, Eve, is assumed to possess full access to both channels. Consequently, all transmitted classical data and quantum states can be perfectly intercepted and stored in her classical and quantum memories, respectively. By granting the adversary direct and complete access to the quantum inputs, we operate under a  threat model analogous to strong quantum data locking \cite{CosmoLupo}.

\paragraph{Time-lock encryption in the security experiment.}
Definition~\ref{time-lock} says that the ciphertext gives Eve the same final information as delayed access to its plaintext. In the simulation, her processing before release is therefore one seed-independent quantum instrument,
\begin{equation}
\mathcal E(\omega_A)=\sum_w |w\rangle\langle w|_W\otimes\mathcal E_w(\omega_A),
\label{eq:time-lock-instrument}
\end{equation}
where $W$ is her classical record, $Q$ is the quantum output, and the completely positive maps $\mathcal E_w$ sum to a trace-preserving map. Its outcomes may depend on the received quantum state. Her retained quantum information then undergoes the effective storage channel $\Phi^Q_{t_{comp}}$ of Definition~\ref{full-decoherence}.

For a signal state $\rho_{x,s}$, conditioned on bit $x$ and seed $s$, the assumed simulation property reads
\begin{equation}
\rho^{x,s}_{E_{rel}}
=\mathcal D_s\!\left[
(\operatorname{id}_W\otimes\Phi^Q_{t_{comp}})
\mathcal E(\rho_{x,s})\right].
\label{eq:time-lock-view}
\end{equation}
Here $E_{rel}$ is Eve's complete information at disclosure, before public post-processing. The same map $\mathcal E$ applies to every $(x,s)$; the channel $\mathcal D_s$ uses the released seed to reconstruct the encryption transcript and perform any subsequent computation, and is independent of $x$. The equality holds simultaneously for all inputs for each admissible Eve, including correlations with an external reference register, on which the maps act as the identity. Outcome-dependent storage is interpreted branchwise in $W$. This equation makes the encryption assumption a property of Eve's full view of the ciphertext and quantum signals. Data processing then transfers the security bounds on $(W,S,Q)$ to her complete view.

Formally, the QCT security model is defined by the following assumptions on the adversary's capabilities:

\begin{Definition}[QCT security model]
\label{QCT-Definition}
The QCT security model rests on two fundamental assumptions:

\begin{enumerate}
\item \textbf{Time-lock encryption:} Alice and Bob use encryption with immediate decryption for Bob and delayed plaintext access for Eve, in the precise simulation sense of Definition~\ref{time-lock}. The protected interval is scheduled as specified in Protocol~1.

\item \textbf{Noisy Quantum Memory:} Every residual quantum register retained by Eve is subject to a $(t_{\mathrm{comp}},\delta)$-noisy quantum memory satisfying Definition~\ref{full-decoherence}.
\end{enumerate}
\end{Definition}
Standard cryptographic primitives such as AES motivate short-term confidentiality when the adversary's computational and parallel resources are bounded \cite{topsecret}; the security analysis assumes the operational time-lock encryption behavior of Definition~\ref{time-lock}. Quantum storage performance depends on platform, encoding and control, including conversion of optical information into other memories \cite{TIon,Tion1,NV,NVC,NVC1,QED,QED1,Raman1,Raman2,Raman3,AFC,AFC1,AFC2,Cavity,EIT,EIT1,GEM,DLCZ}. The parameter $\delta$ quantifies this effective storage constraint. Under the time-lock encryption and storage assumptions, Section~\ref{evemodel} proves the single-round bound

\begin{equation}
\frac{1}{2}
\left\|
\rho_{XWSQ}^{\,t_{comp}}
-
u_X\otimes\rho_{WSQ}^{\,t_{comp}}
\right\|_1
\leq
\frac{1}{2\sqrt{d+1}}+\delta .
\end{equation}
The corresponding asymptotic conditional-entropy bound contains the continuity correction $g(\delta)$ defined in equation~(\ref{continuity-correction}). Thus, if one requires the storage imperfection not to dominate the intrinsic single-round trace-distance term, it is sufficient to operate with $\delta=O(d^{-1/2})$. If one instead requires the memory contribution not to dominate the $\mathcal{O}(1/d)$ entropy penalty relevant to the asymptotic key rate, the stronger condition $g(\delta)=O(1/d)$ is appropriate.

\section{MUB-QCT key establishment protocol}
\label{sec:MUBQCTProtocol}
Following the QCT security model, we now describe a cryptographic primitive for key distribution that we call \textit{MUB-Quantum Computational Timelock} (MUB-QCT).  

\subsection{MUB-QCT Encoding and Decoding}

To construct the MUB-QCT protocol, two ingredients are combined: (a) a complete set of mutually unbiased bases (MUBs) $\{U_{\theta}\}$ for $\theta\in[d+1]$ (Definition \ref{MUBs}), and (b) a family $\mathcal{P}$ of pair-wise independent permutations $\{P_{\sigma} \}$ for $ \sigma \in [|\mathcal{P}|]$ (Definition \ref{permutation}).

Recall that $A$ is a $d$-dimensional quantum system, with $d$ a power of two, associated with the Hilbert space $\mathcal{H}_A = \mathcal{H}_{A_1} \otimes \mathcal{H}_{A_2}$. Let $\{|x\rangle\}_{x\in\{0,1\}}$ and $\{|r\rangle\}_{r\in[d/2]}$ be the standard orthonormal bases of $A_{1}$ and $A_{2}$, respectively. The computational basis elements of the joint system $A$ are defined as $|i_{x,r}\rangle$, where the index is $i_{x,r} = \frac{d}{2}x + r$.

For each $\theta\in[d+1]$, let $\{|e_i^\theta\rangle\}_{i\in[d]}$ denote the corresponding mutually unbiased basis. We fix the unitary convention
\begin{equation}
U_\theta^\dagger |i\rangle = |e_i^\theta\rangle .
\label{MUB-unitary-convention}
\end{equation}
Thus applying $U_\theta$ followed by a computational-basis measurement is equivalent to measuring in MUB $\theta$.

Define the two rank-$d/2$ projectors
\begin{equation}
\Pi_x := \sum_{r\in[d/2]} |i_{x,r}\rangle \langle i_{x,r}|, \qquad x\in\{0,1\}.
\label{binary-projectors}
\end{equation}
They satisfy
\begin{equation}
\Pi_0+\Pi_1=\mathbb{I}_d,
\qquad
\operatorname{Tr}\Pi_x=\frac{d}{2}.
\end{equation}

For a seed $s=(\theta,\sigma)$, define
\begin{equation}
V_{\theta\sigma}
:=
(P_\sigma U_\theta)^\dagger
=
U_\theta^\dagger P_\sigma^\dagger .
\label{encoding-unitary}
\end{equation}
For each channel use Alice samples
$x\in\{0,1\}$, $r\in[d/2]$, $\theta\in[d+1]$, and
$\sigma\in\mathcal P$ uniformly and prepares
\begin{equation}
|\psi_{x,r}^{\theta\sigma}\rangle
:=
V_{\theta\sigma}|i_{x,r}\rangle
=
U_\theta^\dagger P_\sigma^\dagger
|i_{x,r}\rangle .
\label{pureencodedstate}
\end{equation}

Since $P_\sigma^\dagger$ only permutes the computational-basis label,
\begin{equation}
|\psi_{x,r}^{\theta\sigma}\rangle
=
|e_{\sigma^{-1}(i_{x,r})}^{\theta}\rangle ,
\end{equation}
so every transmitted pure state remains a state of MUB $\theta$.

The variable $r$ is private local randomness and is never revealed. Conditioned on $(x,\theta,\sigma)$ but averaged over $r$, the transmitted state is
\begin{equation}
\rho_x^{\theta\sigma}
=
\frac{2}{d}
V_{\theta\sigma}
\Pi_x
V_{\theta\sigma}^{\dagger}
=
\frac{2}{d}
U_\theta^\dagger P_\sigma^\dagger
\Pi_x
P_\sigma U_\theta .
\label{conditionalrhox}
\end{equation}

Averaging additionally over the time-lock-encrypted seed gives
\begin{equation}
\rho_x
=
\frac{1}{(d+1)|\mathcal P|}
\sum_{\theta,\sigma}
\rho_x^{\theta\sigma}.
\label{rhox}
\end{equation}

\noindent The MUB-QCT protocol consists of two phases: 
 
(i) \textit{Time-lock encryption:} Alice encrypts $(\theta,\sigma)$ and Bob decrypts it immediately using his authorized key or trapdoor. The ciphertext has the delayed-access security property against Eve stated in Definition~\ref{time-lock}. With symmetric encryption, Alice and Bob start with shared secret key material and may reuse a short encryption key under the computational assumption. The net key-expansion rate accounts for encryption-key refresh and authentication consumption. Trapdoor time-release encryption supplies the alternative authorized-decryption mechanism described in Appendix~\ref{CryptoDefinitions}.

\begin{figure}[ht]
  \centering
  \includegraphics[width=0.7\linewidth]{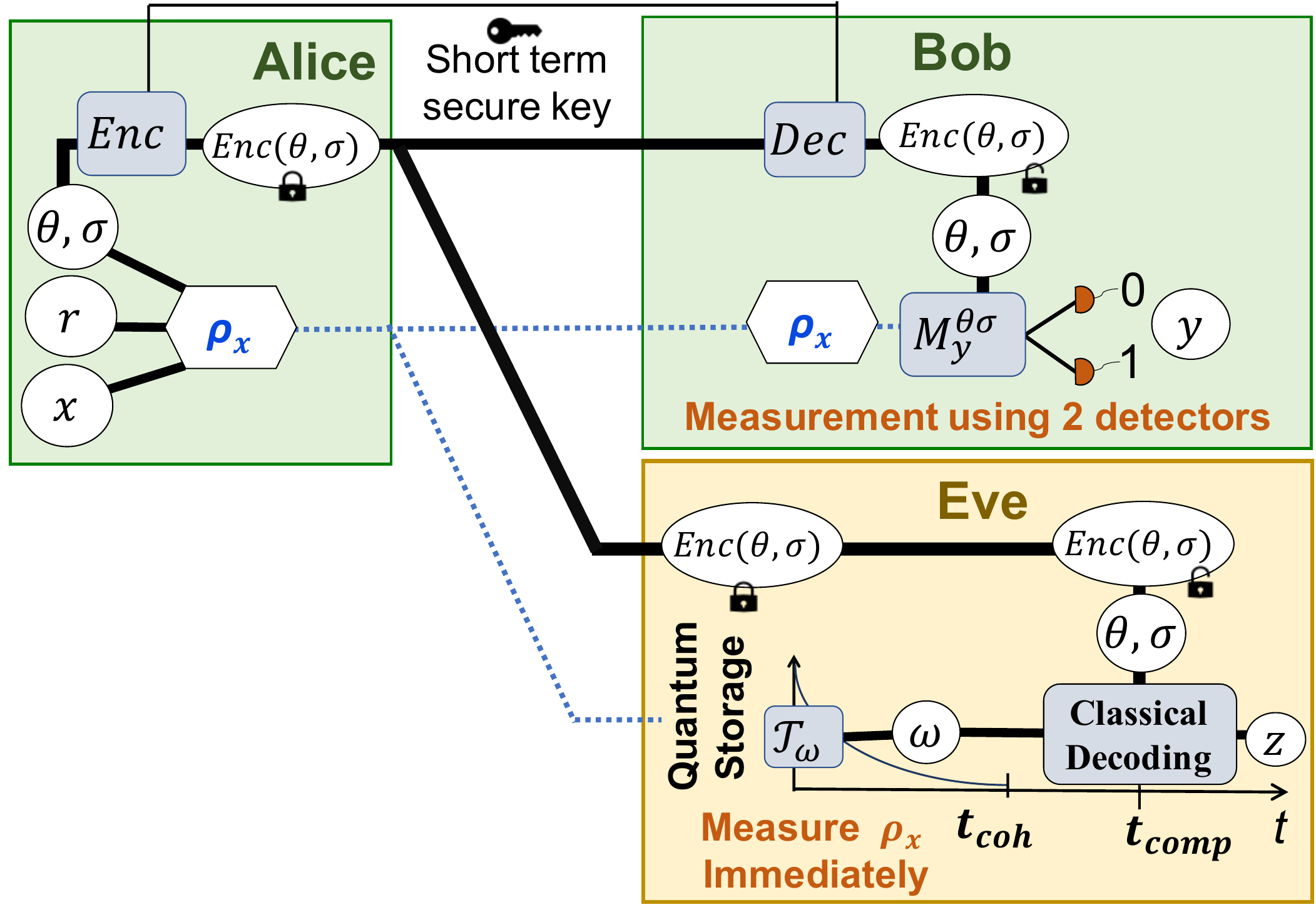}
 \caption{A general overview of MUB-QCT. (a) Communication between the authorized parties. (b) Eve is described by the time-lock encryption assumption of Definition~\ref{time-lock} and the effective noisy-storage bound of Definition~\ref{full-decoherence}. The coherence-time comparison in the diagram is heuristic; security is parameterized by $\delta$. At disclosure Eve learns $(\theta,\sigma)$ and processes her retained information.}
\label{MUBQCT}
\end{figure}

(ii) \textit{Quantum communication:} For each channel use, Alice prepares the pure state $|\psi_{x,r}^{\theta\sigma}\rangle$ of equation~(\ref{pureencodedstate}) using the sampled values $(x,r,\theta,\sigma)$ and sends it to Bob through the insecure quantum channel. The density operator $\rho_x^{\theta\sigma}$ in equation~(\ref{conditionalrhox}) denotes the corresponding state averaged only over Alice's private randomness $r$, while $\rho_x$ in equation~(\ref{rhox}) additionally averages over the time-lock-encrypted seed. After recovering $(\theta,\sigma)$, Bob decodes the bit using the POVM $\{M_y^{\theta\sigma}\}_{y\in\{0,1\}}$:
\begin{equation}
M_y^{\theta\sigma}
=
V_{\theta\sigma}
\Pi_y
V_{\theta\sigma}^\dagger
=
U_\theta^\dagger P_\sigma^\dagger
\Pi_y
P_\sigma U_\theta,
\qquad
y\in\{0,1\}.
\label{measurement}
\end{equation}
Bob's measurement apparatus consists of two detection groups, as shown in Figure \ref{MUBQCT}, a click in a given group corresponding to one of the two bit values $y \in \{0,1\}$. Alice and Bob retain a raw bit pair only for a single logical detection outcome; no-click and double-click events are discarded. The security of this acceptance rule is subject to the acceptance-model assumption below.

\subsection{Key establishment from MUB-QCT}

\begin{Protocol}
\hrule
\vspace{0.1cm}
\textbf{MUB-QCT Key Establishment} 
\hrule
\vspace{0.2cm}
\textit{\textbf{Resources:}}
Alice and Bob have access to the authenticated classical channel, the insecure quantum channel, and the authorized decryption mechanism for the
time-lock encryption of Definition~\ref{time-lock}.

\textit{\textbf{Output:}}
\textbf{Alice:} $S_A\in\{0,1\}^{\ell}$,
\qquad
\textbf{Bob:} $S_B\in\{0,1\}^{\ell}$.

\textbf{\textit{The protocol:}}
\begin{enumerate}
    \item  \textbf{\textit{Data generation:}} Alice chooses $(x^{n},\theta^{n},\sigma^{n},r^{n})$ uniformly at random. 
    
    \item \textbf{\textit{Time-lock encryption:}} Alice and Bob exchange the time-lock-encrypted information $(\theta^{n}, \sigma^{n})$ using the time-lock encryption scheme $(\operatorname{Enc},\operatorname{Dec})$. 
    
    \item \textbf{\textit{Quantum communication:}} 
    \begin{itemize}
        \item For ($i=1$; $i\leq n$; $i++$)
            \begin{itemize}
                \item \textit{State preparation:} Alice prepares $|\psi_{x_i,r_i}^{\theta_i\sigma_i}\rangle$ according to equation~(\ref{pureencodedstate}).
                \item \textit{Distribution:} Alice sends the qudit system over the channel to Bob.
                \item \textit{Measurement:} Bob measures the received system using the POVM $\{M_{y}^{\theta_i\sigma_i}\}_{y\in\{0,1\}}$ of equation~(\ref{measurement}) and records the outcome $y_i$.
            \end{itemize}
    
        \item After $n$ iterations, Alice and Bob output their respective raw strings: $(A: x^{n}; B: y^{n})$.
    \end{itemize}
    
\item \textbf{\textit{Protected interval and timing:}} Let $t_i$ be the latest admissible arrival time for signal $i$, enforced by the authenticated timing schedule. Time-lock encryption protects its seed until at least $t_i+t_{comp}$, and the effective storage bound applies over that interval. For a block, a common disclosure time is chosen no earlier than $\max_i t_i+t_{comp}$. For ciphertexts sent before the block, the encryption protection duration includes the block transmission time. The authenticated schedule accepts signals within their prescribed windows. Acceptance announcements, revealed test data and reconciliation messages are released after the protected interval.

    \item \textbf{\textit{Classical post-processing:}}

    \begin{itemize}
        \item \textit{Parameter estimation:} Alice and Bob publicly reveal a randomly selected subset of their raw data and estimate the QBER, denoted $\widehat p_e$. They abort if $\widehat p_e>p_{\mathrm{th}}$, where $p_{\mathrm{th}}$ is chosen so that the applicable key-rate bound is positive. The remaining accepted raw strings have length $\tilde n$.
        \item \textit{Information reconciliation and privacy amplification:} Alice and Bob perform information reconciliation, followed by two-universal-hash error verification. They then apply a publicly seeded two-universal hash function for privacy amplification and obtain the final keys $S_A$ and $S_B$ of length $\ell$. 
    \end{itemize}
\end{enumerate}
\hrule
\end{Protocol}

\subsection{Eavesdropping model and individual-attack security}
\label{evemodel}

We now establish the single-copy security of MUB-QCT and make explicit the interplay between the two ingredients of the QCT model. Time-lock encryption gives the seed-independent instrument representation before $t_{comp}$, while the noisy-memory assumption bounds the residual quantum information at seed disclosure.

Let the time-lock-encrypted encoding seed be denoted by
 $ S:=(\Theta,\Sigma), $ with realizations $s=(\theta,\sigma)$. The classical outcome of an uninformed measurement performed by Eve before $t_{comp}$ is denoted by $W$, and $Q$ denotes any residual quantum register retained by Eve. Let $E$ be a generic adversarial side information.

We consider individual attacks applied independently across protocol rounds. Eve has complete access to the quantum system transmitted by Alice but has no information about $S$ before $t_{comp}$. At $t_{comp}$ we  reveal $S$ to Eve.

In the per-round proof, time $t_{comp}$ is measured from the start of that signal's protected storage interval. The block schedule includes the transmission duration when setting the expiry of earlier ciphertexts. Effective storage incorporates all controls allowed by the model; outcome-dependent channels are handled branchwise with the same bound $\delta$. The notation suppresses this branch dependence.

The security argument consists of two steps. First, the strong
QC-extractor \cite{QCE} structure bounds the information obtained by any measurement performed before $S$ is known. Second, the noisy-memory condition shows that a general attack retaining both classical and quantum information reduces, up to the memory error $\delta$, to such an immediate measurement.

\paragraph{Binary classical-quantum distance and guessing probability.} We repeatedly use the following identity.

\begin{lemma}[Binary classical-quantum distance]
\label{lem:binary-cq-distance}
Let $X$ be uniformly distributed on $\{0,1\}$ and let
\begin{equation}
\rho_{XE}
=
\frac{1}{2}
\sum_{x=0}^{1}
|x\rangle\langle x|
\otimes
\rho_E^x .
\end{equation}
Then
\begin{equation}
\frac{1}{2}
\left\|
\rho_{XE}
-
u_X\otimes\rho_E
\right\|_1
=
\frac{1}{4}
\left\|
\rho_E^0-\rho_E^1
\right\|_1,
\label{binary-cq-distance}
\end{equation}
where $u_X=\mathbb{I}_2/2$ and
$\rho_E=(\rho_E^0+\rho_E^1)/2$. Moreover,
\begin{equation}
P_{\mathrm{guess}}(X|E)
=
\frac{1}{2}
+
\frac{1}{2}
\left\|
\rho_{XE}
-
u_X\otimes\rho_E
\right\|_1 .
\label{binary-cq-pguess}
\end{equation}
\end{lemma}

\begin{proof}
Equation~(\ref{binary-cq-distance}) follows by writing the difference as a block-diagonal operator in the classical register $X$. The second identity is the Helstrom bound for minimum-error discrimination of two equiprobable quantum states \cite{Helstrom1976}.
\end{proof}

\paragraph{Three attack strategies.}

It is useful to distinguish three strategies.

\begin{enumerate}[I]

\item \textbf{Delayed measurement.}
Eve stores the complete intercepted quantum system without performing a measurement. At $t_{comp}$ she learns $S$ and measures the degraded quantum state using the optimal seed-dependent POVM.

\item \textbf{Immediate measurement.}
Eve performs an arbitrary seed-independent POVM $\{F_w\}_w$ before $t_{comp}$ and stores only its classical outcome $W$. After $t_{comp}$ she learns $S$ and applies arbitrary classical post-processing to $(W,S)$.

\item \textbf{Hybrid measure-and-store attack.}
Eve applies an arbitrary seed-independent quantum instrument $\{\mathcal E_w\}_w$ before $t_{comp}$. She stores its classical outcome $W=w$ and retains the residual quantum register $Q$ in her noisy quantum memory. At $t_{comp}$ she learns $S$ and performs an arbitrary joint measurement on $(W,S,Q)$.

\end{enumerate}

Strategy III contains Strategies I and II as special cases. We analyze it directly: the QCT noisy-memory condition reduces Strategy III, up to an additive error $\delta$, to Strategy II. In the ideal-memory-erasure limit $\delta=0$, Strategy II is therefore optimal among all individual measure-and-store attacks considered here.

\paragraph{\textbf{Strategy I: delayed measurement.}}

We first consider an adversary who stores the complete intercepted system until the seed becomes available. Let $\Phi_{t_{comp}}^{Q}$ denote her physical storage channel. By Definition~\ref{full-decoherence},
\begin{equation}
\frac{1}{2}
\left\|
\Phi_{t_{comp}}^{Q}
-
\mathcal F_Q
\right\|_\diamond
\leq
\delta,
\label{memory-diamond-use}
\end{equation}
where $\mathcal F_Q$ is the completely depolarizing replacement channel.

If $\mathcal F_Q$ acted instead of the real memory, Eve's stored state at $t_{comp}$ would be independent of $X$. Hence the ideal delayed strategy has guessing probability  $1/2$. Using Lemma~\ref{lem:binary-cq-distance} together with equation~(\ref{memory-diamond-use}) gives
\begin{equation}
\frac{1}{2}
\left\|
\rho_{XSQ}^{\operatorname{I}}
-
u_X\otimes\rho_{SQ}^{\operatorname{I}}
\right\|_1
\leq
\delta
\label{delayed-trace-bound}
\end{equation}
and therefore
\begin{equation}
P_{\mathrm{guess}}^{\operatorname{I}}(X|S,Q)
\leq
\frac{1}{2}+\delta .
\label{pguessI}
\end{equation}

\paragraph{\textbf{Strategy II: Immediate Measurement and the QC-extractor duality.}} Let Eve apply an arbitrary seed-independent POVM $\{F_w\}_w$ to the transmitted state. For every outcome satisfying
\begin{equation}
t_w:=\operatorname{Tr}F_w>0,
\end{equation}
define the normalized POVM effect
\begin{equation}
\tau_w
:=
\frac{F_w}{t_w}.
\label{normalized-povm}
\end{equation}
Since $\{F_w\}_w$ is a POVM,
\begin{equation}
\sum_w t_w
=
d.
\label{sum-povm-traces}
\end{equation}
For every seed $(\theta,\sigma)$,
\begin{align}
\frac{1}{2}
\left(
\rho_0^{\theta\sigma}
+
\rho_1^{\theta\sigma}
\right) =
\frac{1}{d}
V_{\theta\sigma}
(\Pi_0+\Pi_1)
V_{\theta\sigma}^{\dagger} =
\frac{\mathbb{I}_d}{d}.
\label{average-state-maximally-mixed}
\end{align}
Therefore
\begin{equation}
p(w|\theta,\sigma)
=
\frac{t_w}{d},
\label{outcome-marginal}
\end{equation}
which is independent of the revealed seed. Using Bayes' rule and
$V_{\theta\sigma}^{\dagger}=P_\sigma U_\theta$, Eve's posterior
distribution after learning $(\theta,\sigma)$ is
\begin{equation}
p(x|w,\theta,\sigma)
=
\operatorname{Tr}
\left[
\Pi_x
P_\sigma U_\theta
\tau_w
U_\theta^\dagger P_\sigma^\dagger
\right].
\label{prepare-measure-duality}
\end{equation}
Equation~(\ref{prepare-measure-duality}) is the key preparation--measurement duality of MUB-QCT. For every fixed outcome $w$, the right-hand side is  the binary measurement distribution generated by applying the QC-extractor \cite{QCE} transformation $P_\sigma U_\theta$ to the source state $\tau_w$ and then applying the measurement map $\mathcal T_{A\rightarrow A_1}$ of \ref{QCExtractor}. Thus the same unitary family that defines Alice's encoding through its adjoint also defines the randomness extractor that limits Eve's uninformed measurement.

\begin{Proposition}[QC-extractor bound for immediate measurement]
\label{prop:QC-immediate}

For every seed-independent POVM $\{F_w\}_w$,
\begin{equation}
\frac{1}{2}
\left\|
\rho_{XWS}
-
u_X\otimes\rho_{WS}
\right\|_1
\leq
\frac{1}{\sqrt{2(d+1)}} .
\label{QC-immediate-bound}
\end{equation}
\end{Proposition}

\begin{proof}
For each POVM outcome $w$, let
$
t_w:=\operatorname{Tr}F_w,
\qquad
p_w:=\frac{t_w}{d},
\qquad
\tau_w:=\frac{F_w}{t_w},
$
where outcomes with $t_w=0$ may be omitted. Define the classical-quantum state
\begin{equation}
\tau_{AW}
:=
\sum_w
p_w\,
\tau_w
\otimes
|w\rangle\langle w|.
\label{dual-cq-source}
\end{equation}
Because $\tau_w\leq\mathbb I_A$ for every normalized density operator,
\begin{equation}
\tau_{AW}
\leq
\mathbb I_A
\otimes
\tau_W,
\qquad
\tau_W
:=
\sum_w p_w|w\rangle\langle w|.
\end{equation}
Hence, by the definition of conditional min-entropy,
\begin{equation}
H_{\min}(A|W)_{\tau}
\geq
0.
\label{dual-source-hmin}
\end{equation}

Equation~(\ref{prepare-measure-duality}) shows that applying $P_\sigma U_\theta$ to the $A$ register of $\tau_{AW}$ and then applying $\mathcal T_{A\rightarrow A_1}$ produces  Eve's joint posterior distribution of $(X,W)$ for the seed $(\theta,\sigma)$.

We may therefore apply Theorem~\ref{QCMUB}, which is the zero-smoothing specialization of Theorem III.8 of Ref.~\cite{QCE}, with
$
|A|=d,
\qquad
|A_1|=2,
\qquad
H_{\min}(A|W)\geq0.
$
This gives the full trace-norm bound
\begin{equation}
\left\|
\rho_{XWS}
-
u_X\otimes\rho_{WS}
\right\|_1
\leq
\sqrt{\frac{2}{d+1}}.
\end{equation}
Dividing by two gives
\begin{equation}
\frac{1}{2}
\left\|
\rho_{XWS}
-
u_X\otimes\rho_{WS}
\right\|_1
\leq
\frac{1}{\sqrt{2(d+1)}}.
\end{equation}
The strong-QC-extractor property is essential because the seed $S$ is revealed after Eve's measurement and is explicitly included in her final side information.
\end{proof}

\paragraph{\textbf{Binary MUB-QCT refinement.}}

The general QC-extractor theorem already bounds every seed-independent immediate POVM. For the binary partition $|A_1|=2$ used by MUB-QCT, we can obtain a sharper constant by specializing the second-moment method underlying the full-MUB QC-extractor proof of Ref.~\cite{QCE}. The calculation uses the pair-wise independence of $\mathcal P$ and the fact that a complete set of MUB vectors forms a complex projective $2$-design
\cite{MUB2design}.

\begin{theorem}[Immediate-measurement security of MUB-QCT]
\label{thm:immediate-measurement}

For the single-copy MUB-QCT encoding and every seed-independent POVM
$\{F_w\}_w$,
\begin{equation}
\frac{1}{2}
\left\|
\rho_{XWS}
-
u_X\otimes\rho_{WS}
\right\|_1
\leq
\frac{1}{2\sqrt{d+1}}.
\label{immediate-trace-bound}
\end{equation}
Consequently,
\begin{equation}
P_{\mathrm{guess}}^{\operatorname{II}}(X|W,S)
\leq
\frac{1}{2}
+
\frac{1}{2\sqrt{d+1}}.
\label{pguessII}
\end{equation}
Moreover,
\begin{equation}
I(X;W,S)
\leq
\frac{1}{d+1}.
\label{immediate-information-bound}
\end{equation}
\end{theorem}

\begin{proof}
Define
\begin{equation}
D
:=
\Pi_0-\Pi_1.
\label{binary-observable}
\end{equation}
Then
\begin{equation}
D^2=\mathbb{I}_d,
\qquad
\operatorname{Tr}D=0,
\end{equation}
and $D$ has $d/2$ eigenvalues $+1$ and $d/2$ eigenvalues $-1$.

For fixed $(w,\theta)$, let
\begin{equation}
q_i^{w,\theta}
:=
\langle i|
U_\theta
\tau_w
U_\theta^\dagger
|i\rangle
=
\langle e_i^\theta|
\tau_w
|e_i^\theta\rangle .
\label{mub-probabilities}
\end{equation}
The coefficients $\{q_i^{w,\theta}\}_{i\in[d]}$ form a probability distribution. Let
\begin{equation}
s_i
:=
\begin{cases}
+1, & 0\leq i<d/2,\\
-1, & d/2\leq i<d.
\end{cases}
\end{equation}
Using equation~(\ref{prepare-measure-duality}), define the posterior bias
\begin{align}
b_{w\theta\sigma}:=
p(0|w,\theta,\sigma)
-
p(1|w,\theta,\sigma)
=
\operatorname{Tr}
\left[
D
P_\sigma U_\theta
\tau_w
U_\theta^\dagger P_\sigma^\dagger
\right]
=
\sum_{i=0}^{d-1}
s_{\sigma(i)}
q_i^{w,\theta}.
\label{posterior-bias}
\end{align}
Pair-wise independence of $\mathcal P$ implies
\begin{equation}
\mathbb{E}_{\sigma}
[s_{\sigma(i)}]
=
0,
\end{equation}
and, for $i\neq j$,
\begin{equation}
\mathbb{E}_{\sigma}
[
s_{\sigma(i)}
s_{\sigma(j)}
]
=
-\frac{1}{d-1}.
\label{pairwise-sign-correlation}
\end{equation}
Therefore
\begin{align}
\mathbb{E}_{\sigma}
[
b_{w\theta\sigma}^{2}
]
=
\sum_i
(q_i^{w,\theta})^2
-
\frac{1}{d-1}
\sum_{i\neq j}
q_i^{w,\theta}q_j^{w,\theta}=
\frac{
d\sum_i(q_i^{w,\theta})^2-1
}{
d-1
}.
\label{permutation-second-moment}
\end{align}

A complete set of $d+1$ mutually unbiased bases forms a complex projective $2$-design \cite{MUB2design}. Hence
\begin{equation}
\sum_{\theta\in[d+1]}
\sum_{i\in[d]}
\left(
|e_i^\theta\rangle
\langle e_i^\theta|
\right)^{\otimes2}
=
\mathbb{I}
+
\mathbb{F},
\label{mub-two-design}
\end{equation}
where $\mathbb{F}$ denotes the swap operator. Taking the expectation in $\tau_w^{\otimes2}$ gives
\begin{equation}
\sum_{\theta\in[d+1]}
\sum_{i\in[d]}
(q_i^{w,\theta})^2
=
1+\operatorname{Tr}(\tau_w^2).
\label{mub-collision-identity}
\end{equation}

Averaging equation~(\ref{permutation-second-moment}) over $\theta$ and using equation~(\ref{mub-collision-identity}) yields
\begin{align}
\mathbb{E}_{\theta,\sigma}
[
b_{w\theta\sigma}^{2}
]
=
\frac{
d\,\operatorname{Tr}(\tau_w^2)-1
}{
d^2-1
}
\leq
\frac{1}{d+1},
\label{bias-second-moment}
\end{align}
where we used
$\operatorname{Tr}(\tau_w^2)\leq1$.

Cauchy--Schwarz gives
\begin{equation}
\mathbb{E}_{\theta,\sigma}
[
|b_{w\theta\sigma}|
]
\leq
\frac{1}{\sqrt{d+1}}.
\label{bias-first-moment}
\end{equation}

For a binary posterior with bias $b$,
\begin{equation}
\frac{1}{2}
\left\|
P_{X|w,\theta,\sigma}
-
u_X
\right\|_1
=
\frac{|b_{w\theta\sigma}|}{2}.
\end{equation}
Using equation~(\ref{outcome-marginal}),
\begin{align}
\frac{1}{2}
\left\|
\rho_{XWS}
-
u_X\otimes\rho_{WS}
\right\|_1
&=
\frac{1}{2d}
\sum_w
t_w
\mathbb{E}_{\theta,\sigma}
[
|b_{w\theta\sigma}|
]
\nonumber\\
&\leq
\frac{1}{2d}
\sum_w
t_w
\frac{1}{\sqrt{d+1}}
\nonumber\\
&=
\frac{1}{2\sqrt{d+1}}.
\end{align}
This proves equation~(\ref{immediate-trace-bound}). The guessing bound (\ref{pguessII}) follows immediately from Lemma~\ref{lem:binary-cq-distance}.

Finally, conditioned on $(w,\theta,\sigma)$,
\begin{equation}
p(0|w,\theta,\sigma)
=
\frac{1+b_{w\theta\sigma}}{2}.
\end{equation}
Thus
\begin{equation}
I(X;W,S)
=
\mathbb{E}_{W,S}
\left[
1-
h_2
\left(
\frac{1+b_{W,S}}{2}
\right)
\right].
\end{equation}
For every $|b|\leq1$,
\begin{equation}
1-
h_2
\left(
\frac{1+b}{2}
\right)
\leq
b^2.
\end{equation}
Using equation~(\ref{bias-second-moment}) and
$\sum_wt_w=d$ therefore gives
\begin{align}
I(X;W,S)
\leq
\frac{1}{d}
\sum_w
t_w
\mathbb{E}_{\theta,\sigma}
[
b_{w\theta\sigma}^{2}
]
\leq
\frac{1}{d+1}.
\end{align}
\end{proof}

Theorem~\ref{thm:immediate-measurement} is therefore a binary prepare-and-measure refinement of the strong-QC-extractor bound (\ref{QC-immediate-bound}). The QC-extractor theorem establishes the security mechanism against arbitrary quantum source states satisfying the appropriate min-entropy condition, while the direct MUB calculation exploits the special two-subspace structure of MUB-QCT to improve the single-round constant.

\paragraph{Tightness of the information bound.}
Measuring in any fixed MUB gives the correct-basis outcome with probability $1/(d+1)$ over the seed; after disclosure it determines $X$ exactly. For every other basis the outcome is independent of $X$. Thus the information bound $1/(d+1)$ is attained. The optimal trace-distance advantage remains a separate optimization question.

\paragraph{\textbf{Strategy III: general hybrid measure-and-store attack.}}

We now consider the most general individual attack allowed by the present QCT model. Before learning $S$, Eve applies an arbitrary quantum instrument
\begin{equation}
\{\mathcal E_w\}_w:
A\longrightarrow WQ.
\end{equation}
The classical outcome $W=w$ is retained without noise, while every residual quantum register is stored in the $(t_{comp},\delta)$-noisy memory. The POVM associated with the instrument is
\begin{equation}
F_w
:=
\mathcal E_w^\dagger(\mathbb{I}_Q),
\end{equation}
so that $\{F_w\}_w$ is precisely a seed-independent immediate POVM on the intercepted system.

The following proposition formalizes the central interplay between the QC-extractor mechanism and the noisy-memory assumption.
\begin{Proposition}[Reduction of hybrid attacks to immediate measurement]
\label{optimalattack}
Let Eve perform an arbitrary seed-independent individual quantuminstrument before $t_{comp}$, retain the classical outcome $W$, and store the residual quantum system $Q$ in a $(t_{comp},\delta)$-noisy memory. Then
\begin{equation}
\frac{1}{2}
\left\|
\rho_{XWSQ}^{\,t_{comp}}
-
u_X\otimes
\rho_{WSQ}^{\,t_{comp}}
\right\|_1
\leq
\Delta_{\mathrm{imm}}(d)
+
\delta,
\label{hybrid-general-bound}
\end{equation}
where
\begin{equation}
\Delta_{\mathrm{imm}}(d)
:=
\sup_{\{F_w\}}
\frac{1}{2}
\left\|
\rho_{XWS}
-
u_X\otimes\rho_{WS}
\right\|_1 .
\label{DeltaImmDefinition}
\end{equation}
By Theorem~\ref{thm:immediate-measurement},
\begin{equation}
\Delta_{\mathrm{imm}}(d)
\leq
\frac{1}{2\sqrt{d+1}},
\end{equation}
and hence
\begin{equation}
\frac{1}{2}
\left\|
\rho_{XWSQ}^{\,t_{comp}}
-
u_X\otimes
\rho_{WSQ}^{\,t_{comp}}
\right\|_1
\leq
\frac{1}{2\sqrt{d+1}}
+
\delta .
\label{hybrid-trace-bound}
\end{equation}
Consequently,
\begin{equation}
P_{\mathrm{guess}}(X|W,S,Q)
\leq
\frac{1}{2}
+
\frac{1}{2\sqrt{d+1}}
+
\delta .
\label{hybrid-pguess}
\end{equation}
Define the optimal hybrid distinguishing advantage by
\begin{equation}
\Delta_{\mathrm{hyb}}(d,\delta)
:=
\sup_{\{\mathcal E_w\}_w}
\frac{1}{2}
\left\|
\rho_{XWSQ}^{\,t_{comp}}
-
u_X\otimes\rho_{WSQ}^{\,t_{comp}}
\right\|_1,
\label{DeltaHybDefinition}
\end{equation}
where the supremum is over all seed-independent instruments allowed by the individual-attack model and whose residual register is stored in a $(t_{comp},\delta)$-noisy memory.

For an ideal memory erasure, $\delta=0$, the optimal hybrid individual attack has  the same maximum distinguishing advantage as the optimal immediate-measurement attack:
\begin{equation}
\Delta_{\mathrm{hyb}}(d,0)
=
\Delta_{\mathrm{imm}}(d).
\label{ideal-optimality}
\end{equation}
Thus immediate measurement is optimal in the ideal-erasure limit. For finite $\delta$, a hybrid attack can improve upon the optimal immediate measurement by at most $\delta$.
\end{Proposition}

\begin{proof}
Let $\omega_{WSQ}^{x}$, denote Eve's state conditioned on $X=x$ immediately after her seed-independent instrument and with the seed register $S$ included. The real state at $t_{comp}$ is 
\begin{equation}
\rho_{WSQ}^{x}
=
\left(
\operatorname{id}_{WS}
\otimes
\Phi_{t_{comp}}^{Q}
\right)
\left(
\omega_{WSQ}^{x}
\right).
\end{equation}
Introduce the ideal state obtained by replacing Eve's memory by
$\mathcal F_Q$:
\begin{equation}
\widetilde{\rho}_{WSQ}^{x}
=
\left(
\operatorname{id}_{WS}
\otimes
\mathcal F_Q
\right)
\left(
\omega_{WSQ}^{x}
\right).
\end{equation}
Since $\mathcal F_Q$ replaces $Q$ by a fixed maximally mixed state, the register $Q$ in $\widetilde{\rho}$ is independent of $X$. The only information surviving until the seed reveal is therefore the classical instrument outcome $W$. Moreover, the distribution of $W$ is  that generated by the POVM
\begin{equation}
F_w
=
\mathcal E_w^\dagger(\mathbb{I}_Q).
\end{equation}
Hence the ideal hybrid attack is equivalent to an immediate seed-independent POVM followed by disclosure of $S$. By Definition~(\ref{DeltaImmDefinition}),
\begin{equation}
\frac{1}{4}
\left\|
\widetilde{\rho}_{WSQ}^{0}
-
\widetilde{\rho}_{WSQ}^{1}
\right\|_1
\leq
\Delta_{\mathrm{imm}}(d).
\label{ideal-hybrid-immediate}
\end{equation}

For the real attack,
\begin{align}
\frac{1}{4}
\left\|
\rho_{WSQ}^{0}
-
\rho_{WSQ}^{1}
\right\|_1
&\leq
\frac{1}{4}
\left\|
\widetilde{\rho}_{WSQ}^{0}
-
\widetilde{\rho}_{WSQ}^{1}
\right\|_1
\nonumber\\
&\quad
+
\frac{1}{4}
\left\|
\left[
\operatorname{id}_{WS}
\otimes
(\Phi_{t_{comp}}^{Q}-\mathcal F_Q)
\right]
\left(
\omega_{WSQ}^{0}
-
\omega_{WSQ}^{1}
\right)
\right\|_1 .
\label{hybrid-distance-step}
\end{align}
By the definition of the diamond norm,
\begin{align}
&
\frac{1}{4}
\left\|
\left[
\operatorname{id}_{WS}
\otimes
(\Phi_{t_{comp}}^{Q}-\mathcal F_Q)
\right]
\left(
\omega_{WSQ}^{0}
-
\omega_{WSQ}^{1}
\right)
\right\|_1
\nonumber\\
&\qquad\leq
\frac{1}{4}
\left\|
\Phi_{t_{comp}}^{Q}-\mathcal F_Q
\right\|_\diamond
\left\|
\omega_{WSQ}^{0}
-
\omega_{WSQ}^{1}
\right\|_1
\nonumber\\
&\qquad\leq
\frac{1}{4}
(2\delta)(2)
=
\delta .
\label{hybrid-memory-correction}
\end{align}
Combining equations~(\ref{ideal-hybrid-immediate}) and (\ref{hybrid-memory-correction}), and using Lemma~\ref{lem:binary-cq-distance}, gives equation~(\ref{hybrid-general-bound}). Applying Theorem~\ref{thm:immediate-measurement} proves equations~(\ref{hybrid-trace-bound}) and
(\ref{hybrid-pguess}).

Finally, for $\delta=0$, every hybrid attack is reduced  to its classical POVM outcome $W$. Conversely, every immediate POVM is itself a special case of a hybrid instrument with no retained quantum register. The two optimizations therefore coincide, proving equation~(\ref{ideal-optimality}).
\end{proof}

\paragraph{\textbf{Interpretation of the optimal strategy.}}

The QCT model therefore leads to a precise form of optimality. If Eve's memory is completely depolarized by the time the time-lock-encrypted seed is revealed, retaining a quantum system provides no advantage: every individual hybrid attack is  equivalent to an immediate seed-independent measurement followed by classical decoding once $(\theta,\sigma)$ becomes available. Immediate measurement is therefore the optimal individual attack for $\delta=0$.

For a realistic $(t_{comp},\delta)$-noisy memory, the surviving quantum register can improve Eve's distinguishing advantage by at most $\delta$. Thus the security parameter separates naturally into 
\begin{equation}
\underbrace{\Delta_{\mathrm{imm}}(d)}_{\text{uninformed measurement}}
+
\underbrace{\delta}_{\text{residual quantum memory}},
\end{equation}
with
\begin{equation}
\Delta_{\mathrm{imm}}(d)
\leq
\frac{1}{2\sqrt{d+1}}.
\end{equation}
The first contribution is controlled by the MUB/QC-extractor structure; the second is controlled by the physical QCT memory assumption.

\paragraph{\textbf{Entropy bound for asymptotic key generation.}}

The trace-distance bound above is appropriate for composable distinguishability. For asymptotic key generation, the relevant quantity is the conditional von Neumann entropy.

In the ideal depolarized branch, $Q$ is independent of $X$ and Theorem~\ref{thm:immediate-measurement} gives
\begin{equation}
I(X;W,S)_{\widetilde{\rho}}
\leq
\frac{1}{d+1}.
\end{equation}
Since $X$ is uniform,
\begin{equation}
H(X|W,S,Q)_{\widetilde{\rho}}
=
H(X|W,S)_{\widetilde{\rho}}
\geq
1-\frac{1}{d+1}.
\label{ideal-conditional-entropy}
\end{equation}

The real and ideal hybrid states satisfy
\begin{equation}
\frac{1}{2}
\left\|
\rho_{XWSQ}
-
\widetilde{\rho}_{XWSQ}
\right\|_1
\leq
\delta .
\end{equation}
Define
\begin{equation}
g(\delta)
:=
2\delta
+
(1+\delta)
h_2
\left(
\frac{\delta}{1+\delta}
\right).
\label{continuity-correction}
\end{equation}
The uniform continuity bound for conditional entropy \cite{WinterContinuity} then gives
\begin{equation}
H(X|W,S,Q)_{\rho}
\geq
1
-
\frac{1}{d+1}
-
g(\delta).
\label{hybrid-conditional-entropy}
\end{equation}
Thus the single-round trace-distance deviation from an ideal uniform bit scales as
$
\mathcal O(d^{-1/2})+\delta,
$
for single-round composable distinguishability, while
\begin{equation}
1-H(X|W,S,Q)
=
\mathcal O(d^{-1})+g(\delta)
\end{equation}
for the asymptotic entropy penalty entering the Devetak--Winter key rate \cite{DevetakWinter}. These statements are compatible because trace distance and mutual information quantify different aspects of Eve's side information.
\subsection{Security of MUB-QCT key establishment protocol}
\label{security}

The security of MUB-QCT consists of correctness and secrecy. Let $\Omega$ denote the event that the protocol does not abort, let $S_A$ and $S_B$ be the final keys, and let $E_{\mathrm{fin}}$ denote Eve's complete side information after expiry of the time-lock encryption. The register $E_{\mathrm{fin}}$ includes all classical measurement outcomes, the revealed seed $(\theta,\sigma)$, the public post-processing transcript, and any residual quantum register.

Following the standard composable security criterion for quantum key distribution \cite{RennerPhD,Renner}, we use the following definition.

\begin{Definition}[Composable security]
\label{Sec-Def}
The MUB-QCT protocol is $(\varepsilon_{cor},\varepsilon_{sec})$-secure if the following conditions hold.

\begin{enumerate}[(i)]
\item \textbf{\textit{Correctness:}}
\begin{equation}
\Pr[S_A\neq S_B\wedge\Omega]
\leq
\varepsilon_{cor}.
\label{correctness}
\end{equation}

\item \textbf{\textit{Secrecy:}} writing $p_{\mathrm{pass}}:=\Pr[\Omega]$ and $\tau_{S_A}:=\mathbb{I}_{S_A}/2^\ell$,
\begin{equation}
p_{\mathrm{pass}}\,
\frac{1}{2}
\left\|
\rho_{S_AE_{\mathrm{fin}}\mid\Omega}
-
\tau_{S_A}\otimes
\rho_{E_{\mathrm{fin}}\mid\Omega}
\right\|_1
\leq
\varepsilon_{sec}.
\label{composable-secrecy}
\end{equation}
\end{enumerate}

The total distinguishing advantage from an ideal key resource is at most $\varepsilon_{cor}+\varepsilon_{sec}$.
\end{Definition}

Because equation~(\ref{composable-secrecy}) is an information-theoretic
trace-distance statement evaluated after the time-lock encryption has expired, allowing
Eve unlimited computation afterwards cannot increase the distinguishing
advantage: trace distance is contractive under subsequent quantum
operations. This is the sense in which the resulting key has everlasting
secrecy within the QCT threat model.

\begin{lemma}[Privacy amplification with explicit pass conditioning]
\label{lem:privacy-amplification}
Let $\omega^\Omega_{X^{\tilde n}E_{pre}}$ be the subnormalized state on the final non-abort branch, of trace $p_{pass}$, before the reconciliation and verification transcript is appended to Eve's information. Here $E_{pre}$ includes the revealed encoding seeds, accepted positions, parameter-estimation data and other prior public information. Smooth min-entropy is defined using purified distance over subnormalized states. Let $\operatorname{leak}_{EC}$ bound the length of the reconciliation and verification transcript, including data-dependent public metadata. If
\begin{equation}
\ell\leq H_{\min}^{\varepsilon_s}(X^{\tilde n}|E_{pre})_{\omega^\Omega}
-\operatorname{leak}_{EC}
-2\log_2\!\left(\frac1{2\varepsilon_{PA}}\right),
\label{PA-key-length}
\end{equation}
then an independent public two-universal hash yields secrecy in the actual-marginal criterion~(\ref{composable-secrecy}) with
\begin{equation}
\varepsilon_{sec}\leq2\varepsilon_s+\varepsilon_{PA}.
\end{equation}
The hash seed is included in $E_{fin}$.
\end{lemma}
\begin{proof}
The min-entropy chain rule subtracts the bounded transcript length. The actual-marginal form of the quantum leftover-hash bound gives
\begin{equation}
D\leq2\varepsilon_s+
\frac12\sqrt{2^{\ell-H_{\min}^{\varepsilon_s}(X^{\tilde n}|E_{pre})_{\omega^\Omega}
+\operatorname{leak}_{EC}}}.
\end{equation}
Here $D$ is the subnormalized trace distance appearing in~(\ref{composable-secrecy}). Substitution proves the claim. The distance-to-uniform convention in Ref.~\cite{LeftoverHash} can optimize over an auxiliary marginal; the conservative smoothing term used here accounts for comparison with Eve's actual marginal. Application of the lemma uses a lower bound on the min-entropy of the specified pass-restricted state.
\end{proof}

Under the individual i.i.d.\ attack model of Section~\ref{evemodel}, the
single-round conditional entropy satisfies
\begin{equation}
H(X|E_{\mathrm{round}})
\geq
1-\frac{1}{d+1}-g(\delta),
\label{single-round-entropy}
\end{equation}
where
$E_{\mathrm{round}}=(W,S,Q)$ denotes Eve's complete side information for
one round after expiry of the time-lock encryption.

\paragraph{Acceptance-model assumption.}
\label{ass:acceptance}
For the asymptotic rates, we assume that receiver acceptance preserves the single-round conditional-entropy bound and the i.i.d. structure of the retained rounds. We use independent thinning of the full joint state $(X,S,W,Q)$, conditioned on fixed physical parameters, as an explicit sufficient acceptance model. The detection formulas in Appendices~\ref{CSKR} and~\ref{App:MultiCopy} describe the resulting performance. Adversarially correlated acceptance is addressed by a direct bound on the accepted, pass-conditioned raw-key entropy.

For $\tilde n$ independent accepted rounds in the i.i.d.\ model, the fully
quantum asymptotic equipartition property
\cite{FullyQuantumAEP} gives
\begin{equation}
H_{\min}^{\varepsilon_s}
\left(
X^{\tilde n}\middle|E_{\mathrm{round}}^{\tilde n}
\right)
\geq
\tilde n
\left[
1-\frac{1}{d+1}-g(\delta)
\right]
-
O(\sqrt{\tilde n}),
\label{AEP-raw-key}
\end{equation}

The entropy in~(\ref{AEP-raw-key}) is evaluated on the i.i.d. accepted-round state before finite-sample parameter-estimation abort conditioning and the post-processing transcript. The sublinear correction depends on the smoothing parameter and vanishes per bit in the asymptotic limit. Together with asymptotically negligible sampling costs and the reconciliation leakage, this establishes the conditional asymptotic rate below. The finite-key quantity is the pass-restricted entropy in Lemma~\ref{lem:privacy-amplification}.

A finite-key extension includes explicit parameter-estimation and sampling terms, pass conditioning, authentication accounting and transcript-length bounds. Parameter-estimation information is accounted for once in $E_{pre}$. The source-level Eve-information bound is independent of a microscopic model of Bob's POVM; using it for accepted key generation still requires the stated receiver, laboratory and postselection assumptions.

\begin{lemma}[MUB-QCT: Correctness]
The MUB-QCT protocol is $\varepsilon_{cor}$-correct with
\begin{equation}
\varepsilon_{cor}=2^{-\ell_{hash}},
\end{equation}
where $\ell_{hash}$ is the length of the error-verification tag.
\end{lemma}

\begin{proof}
After information reconciliation, Alice and Bob evaluate a 2-universal
hash function \cite{EC} on their respective reconciled strings and compare
the resulting $\ell_{hash}$-bit verification tags. If the tags differ,
they abort. By 2-universality, two distinct reconciled strings produce the
same verification tag with probability at most $2^{-\ell_{hash}}$.
Consequently,
\begin{equation}
\Pr[S_A\neq S_B\wedge\Omega]
\leq
2^{-\ell_{hash}}
=
\varepsilon_{cor}.
\end{equation}
\end{proof}

\subsection{Asymptotic Secret Key Rate}
\label{sec:AsymptoticKeyRate}

We now state the asymptotic secret fraction under the following assumptions: time-lock encryption as in Definition~\ref{time-lock}, the effective per-round storage bound, the protected timing schedule, i.i.d. individual attacks and entropy-preserving acceptance.

Let $E_{\mathrm{round}}=(W,S,Q)$ denote Eve's complete
classical--quantum side information for one accepted channel use after
expiry of the time-lock encryption. Under the individual i.i.d.\ attack model and the acceptance-model assumption stated above, the asymptotic Devetak--Winter bound \cite{DevetakWinter} gives
\begin{equation}
r
\geq
H(X|E_{\mathrm{round}})
-
H(X|Y).
\end{equation}
Using equation~(\ref{hybrid-conditional-entropy}) and, for binary $X$,
\begin{equation}
H(X|Y)\leq h_2(p_e),
\end{equation}
we obtain
\begin{equation}
r
\geq
1
-
\frac{1}{d+1}
-
g(\delta)
-
h_2(p_e).
\label{single-copy-secret-fraction}
\end{equation}

Writing $P_{\mathrm{click}}$ for the probability that the channel use is
accepted, the asymptotic secret key rate per channel use is therefore
\begin{equation}
R
\geq
P_{\mathrm{click}}
\left[
1
-
\frac{1}{d+1}
-
g(\delta)
-
h_2(p_e)
\right].
\label{single-copy-key-rate}
\end{equation}

The usable rate is truncated at zero. Equation~(\ref{single-copy-key-rate}) is a gross asymptotic rate with ideal reconciliation; an implementation substitutes its actual reconciliation leakage and deducts consumed authentication and encryption-refresh resources to obtain a net expansion rate.

The dimension-dependent secrecy penalty is thus 
$1/(d+1)=\mathcal{O}(1/d)$ in this asymptotic individual-attack model.

\subsection{Analysis}

\textbf{Noise tolerance:}
The immediate-measurement information penalty is bounded by
$1/(d+1)=\mathcal{O}(1/d)$. From
equation~(\ref{single-copy-secret-fraction}), a positive asymptotic secret
fraction is obtained whenever
\begin{equation}
h_2(p_e)
<
1-\frac{1}{d+1}-g(\delta).
\end{equation}
Consequently, for sufficiently small $\delta$, the tolerable QBER
approaches $50\%$ as $d\rightarrow\infty$.

\noindent
\textbf{Resource requirements:}
In conventional prepare-and-measure high-dimensional QKD, a receiver that fully resolves all $d$ outcomes requires $d$ distinguishable detection channels, which can lead to hardware complexity that scales with the encoding dimension \cite{HD1,HD2,HD3,HD4}. In MUB-QCT, Bob does not need to resolve all $d$ basis outcomes individually. After applying the inverse encoding transformation, he only needs to distinguish the two rank-$d/2$ logical subspaces associated with the binary outcomes $y\in\{0,1\}$. The receiver can therefore be implemented with two logical detection outcomes, independently of $d$, as illustrated in Figure~\ref{implementation}. This separates the high-dimensional encoding space from the final binary readout and reduces the detector-resolution requirements of the receiver.

The two logical outcomes simplify the final readout. The total implementation resource model also includes the basis transformation, mode routing, optical loss and the number of physical time or spatial modes.

\noindent
\textbf{Receiver-side modeling.}
The bound on Eve's information derived in
Section~\ref{evemodel} depends on Alice's characterized source ensemble and
the QCT adversarial constraints, but not on a detailed microscopic model of
Bob's POVM. This reduces the receiver-side characterization required by the
secrecy proof. It should not, however, be interpreted as a standard
measurement-device-independent QKD security proof: the analysis still
assumes trusted classical post-processing, controlled postselection, no
unauthorized information leakage from Bob's laboratory, and the receiver
statistics specified by the implementation model.

\begin{center}
\begin{table}
\caption{Comparison of prepare-and-measure HD-QKD with the MUB-QCT protocol.}
\small\renewcommand{\arraystretch}{1.12}\begin{tabularx}{\linewidth}{@{}>{\raggedright\arraybackslash}p{2.1cm}>{\raggedright\arraybackslash}p{1.4cm}>{\raggedright\arraybackslash}p{2.2cm}>{\raggedright\arraybackslash}p{1.7cm}>{\raggedright\arraybackslash}X@{}}
\br
Protocol & Bits/use & QBER limit & Logical outcomes & Trust assumptions \\
\mr
P\&M HD-QKD & $\log_2d$ & $\to50\%$ as $d\to\infty$ & $d$ & Source and receiver devices are characterized. \\
MUB-QCT & $1$ & $\to50\%$ for large $d$, small $\delta$ & $2$ & Time-lock encryption, effective storage constraint, trusted laboratories and entropy-preserving acceptance. \\
\br
\end{tabularx}

\label{Comparison_table}
\end{table}
 \end{center}

 \section{MUB-QCT with multiple independently randomized signal states}
\label{GuessingProbabilitym}

The single-copy security theorem applies to $m=1$. To improve the detection
probability in lossy channels, we consider a multi-copy variant in which
Alice encodes the same logical bit $X$ using the same seed
$S=(\Theta,\Sigma)$ but samples independent private randomizers
$R_1,\ldots,R_m$. She therefore transmits
\begin{equation}
\rho_{x,s}^{(m)}
=
\left(
\rho_x^s
\right)^{\otimes m}.
\label{multicopy-state}
\end{equation}
Copy $j$ is prepared as
$|\psi_{x,r_j}^{s}\rangle$ with an independently sampled $r_j$.

Repeating the bit and seed creates a distinction between separate and joint measurements. We first determine the exact leakage for non-adaptive product measurements, then construct a logarithmic-copy joint attack in Section~\ref{sec:joint-attack}. Equation~(\ref{multicopy-state}) makes the shared-bit and shared-seed correlations explicit despite independent local randomization. Product measurement is an additional operational restriction used for the performance analysis, and the joint-measurement theorem quantifies the information accessible in the broader measurement class.

\subsection{Security analysis in restricted scenarios}

In this section we restrict the multi-copy analysis to attacks in which Eve
performs product measurements before $t_{comp}$ and retains only classical
measurement outcomes. Since the time-lock-encrypted seed
$S=(\Theta,\Sigma)$ is unavailable during these measurements, each POVM
must be chosen independently of the actual encoding seed. This restriction
excludes residual quantum storage, joint coherent measurements across the
$m$ copies, and sequential adaptation of later POVMs to earlier outcomes
unless explicitly stated. Section~\ref{sec:joint-attack} gives a joint-measurement attack within a repeated-signal block. Positive entropy bounds at smaller copy counts and coherent attacks across independently seeded rounds are further research questions.

\subsubsection{Fixed non-adaptive product measurement}

Let $Z_j$ denote the outcome obtained from copy $j$, and write
$Z^m:=(Z_1,\ldots,Z_m)$. Because Alice samples the private randomizers
$R_1,\ldots,R_m$ independently, the outcomes $Z_1,\ldots,Z_m$ are
conditionally independent given $(X,S)$.
Since $X$ and the seed $S$ are independent,
\begin{equation}
I(X;Z_1,\ldots,Z_m,S)
=
I(X;Z_1,\ldots,Z_m|S).
\end{equation}
Using subadditivity of conditional entropy and conditional independence
given $(X,S)$,
\begin{align}
I(X;Z_1,\ldots,Z_m|S)
&=
H(Z_1,\ldots,Z_m|S)
-
H(Z_1,\ldots,Z_m|X,S)
\nonumber\\
&\leq
\sum_{i=1}^{m}
\left[
H(Z_i|S)-H(Z_i|X,S)
\right]
\nonumber\\
&=
\sum_{i=1}^{m}
I(X;Z_i|S).
\end{align}
Theorem~\ref{thm:immediate-measurement} gives
\begin{equation}
I(X;Z_i,S)
=
I(X;Z_i|S)
\leq
\frac{1}{d+1}
\end{equation}
for every copy. Therefore
\begin{equation}
I^{\operatorname{Fix}}
(X;Z_1,\ldots,Z_m,S)
\leq
\min\left\{
1,\frac{m}{d+1}
\right\}.
\label{fixed-multicopy-information}
\end{equation}

For $m\ll d$, Eve's information in this restricted attack model therefore
scales as
\begin{equation}
I^{\operatorname{Fix}}
(X;Z_1,\ldots,Z_m,S)
=
\mathcal{O}\left(\frac{m}{d}\right).
\end{equation}

\subsubsection{Proactive MUB measurement}

We now consider a particular non-adaptive product strategy in which Eve
measures the $m$ copies in distinct bases chosen from the complete set of
$d+1$ MUBs. Let
\begin{equation}
\mu_m
:=
\min\left\{
1,\frac{m}{d+1}
\right\}.
\label{mu-m}
\end{equation}

For the encoding convention used throughout this work,
\begin{equation}
V_{\theta\sigma}
=
U_\theta^\dagger P_\sigma^\dagger,
\end{equation}
and therefore
\begin{equation}
V_{\theta\sigma}|i\rangle
=
|e_{\sigma^{-1}(i)}^\theta\rangle.
\end{equation}
If Eve measures one copy in MUB $\theta$, her measurement outcome is
$\sigma^{-1}(i_{x,r})$. After the seed is revealed, she applies $\sigma$ to
the observed label and recovers $i_{x,r}$, and hence $x$, exactly.

If she measures in a different MUB $\theta'\neq\theta$, mutual unbiasedness
gives
\begin{equation}
\left|
\langle e_j^{\theta'}|
e_{\sigma^{-1}(i_{x,r})}^{\theta}
\rangle
\right|^2
=
\frac{1}{d},
\end{equation}
so the wrong-basis outcome is independent of $x$.

For $m\leq d+1$, choosing $m$ distinct MUBs therefore includes the correct
basis with probability $m/(d+1)$. For $m\geq d+1$, Eve can include every
basis and certainly measures at least one copy in the correct basis.
Consequently,
\begin{equation}
P_{\mathrm{guess}}^{\operatorname{Pro}}
(X|Z,S)
=
\mu_m
+
(1-\mu_m)\frac{1}{2}
=
\frac{1}{2}
+
\frac{\mu_m}{2},
\label{proactive-pguess}
\end{equation}
and the corresponding mutual information is
\begin{equation}
I^{\operatorname{Pro}}(X;Z,S)
=
\mu_m
=
\min\left\{
1,\frac{m}{d+1}
\right\}.
\label{proactive-information}
\end{equation}

Combining equations~(\ref{fixed-multicopy-information}) and~(\ref{proactive-information}) gives the exact restricted optimum,
\begin{equation}
\sup_{\text{non-adaptive product POVMs}} I(X;Z^m,S)
=\min\{1,m/(d+1)\}.
\end{equation}
An independently randomized public measurement schedule can be included in the record without changing this conclusion. In particular, $m=d+1$ gives exact recovery in this restricted class.

\subsection{A logarithmic-copy joint-measurement attack}
\label{sec:joint-attack}

Let $\mathcal A$ be the set of distinct subsets of computational labels
\[
 \mathcal A=\left\{\sigma^{-1}
 \bigl(\{dx/2,\ldots,dx/2+d/2-1\}\bigr):
 x\in\{0,1\},\ \sigma\in\mathcal P\right\},
 \qquad K=|\mathcal A|\leq2|\mathcal P|.
\]
For $A\in\mathcal A$, write
\[
 P_{\theta,A}=\sum_{i\in A}|e_i^\theta\rangle\langle e_i^\theta|,
 \qquad \rho_{\theta,A}=\frac2d P_{\theta,A}.
\]
The independently randomized encoding of a shared bit and seed is
$\rho_{\theta,A}^{\otimes m}$ for the corresponding subset $A$.
All logarithms below are base two.

\begin{theorem}[Joint-measurement raw-bit recovery]
\label{thm:joint-attack}
Suppose the adversary may perform an arbitrary seed-independent joint
measurement on $m=k+1$ signals before the seed is disclosed, retaining
only classical information afterwards. There exists such an attack with
\[
 p_{\mathrm{err}}\leq\min\left\{\frac12,\,2dK2^{-k}\right\}.
\]
Consequently, for $0<\epsilon<1/2$, an error at most $\epsilon$ is
achievable with
\[
 m=1+\left\lceil\log_2\frac{2dK}{\epsilon}\right\rceil.
\]
For the affine family used throughout this work, $|\mathcal P|=d(d-1)$, it is sufficient to take
\[
 m=1+\left\lceil\log_2\frac{4d^2(d-1)}{\epsilon}\right\rceil
 =3\log_2d+\log_2(1/\epsilon)+O(1).
\]
\end{theorem}

\begin{proof}
Reserve the last signal. For each candidate basis define the orthogonal
projector onto the span of all possible $k$-signal supports,
\[
 Q_\theta=\operatorname{proj}\left(
 \sum_{A\in\mathcal A}\operatorname{ran}
 P_{\theta,A}^{\otimes k}\right).
\]
For fixed $\theta$, the projectors in the sum commute, being diagonal
in the same product basis. Thus
\[
 Q_\theta\leq\sum_{A\in\mathcal A}P_{\theta,A}^{\otimes k}.
\]
Let the true basis and subset be $(\theta_0,A_0)$ and write
$\omega=\rho_{\theta_0,A_0}^{\otimes k}$.
Then $\operatorname{Tr}(Q_{\theta_0}\omega)=1$.
For every $\theta\ne\theta_0$ and every $A\in\mathcal A$,
mutual unbiasedness gives
\[
 \operatorname{Tr}(P_{\theta,A}\rho_{\theta_0,A_0})
 =\frac2d\sum_{i\in A,j\in A_0}
 |\langle e_i^\theta|e_j^{\theta_0}\rangle|^2
 =\frac12.
\]
It follows that $\operatorname{Tr}(Q_\theta\omega)\leq K2^{-k}$.

Set $G=\sum_\theta Q_\theta$ and define
$M_\theta=G^{-1/2}Q_\theta G^{-1/2}$, where the inverse is on the
support of $G$. Complete this to a POVM with
$I-\operatorname{proj}(\operatorname{supp}G)$, which has probability
zero for every possible true state. The Hayashi--Nagaoka inequality
\cite[Lemma 2]{HayashiNagaoka2003}, with parameter $c=1$, implies
\[
 I-M_{\theta_0}\leq
 2(I-Q_{\theta_0})+4\sum_{\theta\ne\theta_0}Q_\theta.
\]
Therefore, pointwise for every true bit and seed,
\[
 \Pr(\widehat\Theta\ne\theta_0\mid x,\theta_0,\sigma)
 \leq 4dK2^{-k}.
\]

Eve measures the reserved signal in basis $\widehat\Theta$ and records
the resulting label $J$. After disclosure of $(\Theta,\Sigma)$ she
outputs $\widehat X=\lfloor2\Sigma(J)/d\rfloor$.
If $\widehat\Theta=\Theta$ the answer is exact. Otherwise $J$ is uniform,
by mutual unbiasedness and conditional independence of the reserved
signal from the first $k$ signals. Hence its decoded bit is incorrect
with probability $1/2$, giving
\[
 p_{\mathrm{err}}=\tfrac12\Pr(\widehat\Theta\ne\Theta)
 \leq 2dK2^{-k}.
\]
The measurement can be viewed as a single seed-independent POVM with
effects $M_\theta\otimes|e_j^\theta\rangle\langle e_j^\theta|$.
All its quantum operations occur before disclosure.
\end{proof}

\paragraph{Linear field labeling.}
If the labels identify $\mathrm{GF}(2^q)$ with $\mathbb F_2^q$ linearly,
and the encoded bit is a fixed nonzero linear coordinate $L$, the
subsets have the form $\{i:L(ai+b)=x\}$.
As $a\ne0$ varies, $L(ai)$ runs over all $d-1$ nonzero linear
functionals. Thus $K=2(d-1)$, and the sufficient copy count improves to
\[
 m=1+\left\lceil\log_2\frac{4d(d-1)}{\epsilon}\right\rceil
 =2\log_2d+\log_2(1/\epsilon)+O(1).
\]
The improved constant applies to this linear field-label convention; the general theorem applies to any labeling.

\paragraph{Interpretation.}
For the resulting classical record $W$, an error at most $\epsilon<1/2$
implies
\[
 P_{\mathrm{guess}}(X|W,S)\geq1-\epsilon,
 \qquad H(X|W,S)\leq h_2(\epsilon),
\]
\[
 D(\rho_{XWS},u_X\otimes\rho_{WS})\geq\tfrac12-\epsilon.
\]
The logarithmic count is an upper bound on the copies sufficient for raw-bit recovery. Matching lower bounds and positive security at smaller copy counts remain open. The attack gives substantially larger leakage than the product-measurement optimum, and retains only classical information until disclosure. Residual entropy at finite copy counts can be considered separately for privacy amplification.

\begin{figure}[ht]
  \centering
  \includegraphics[width=\linewidth]{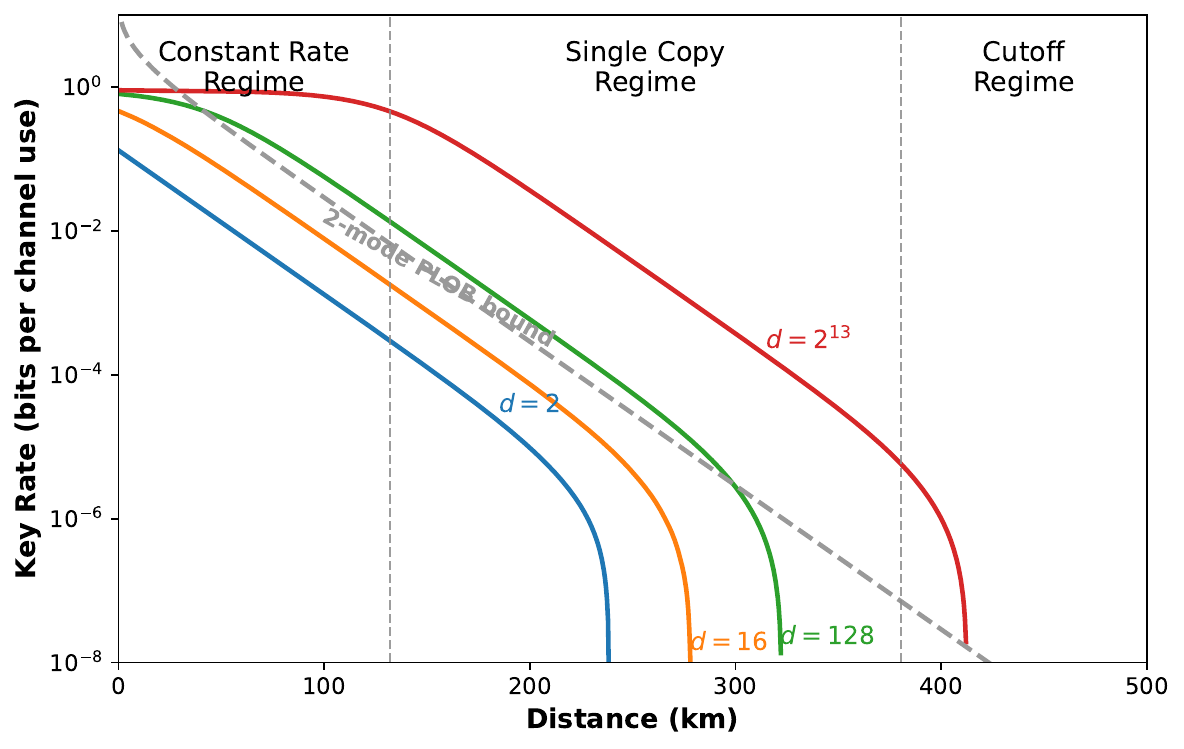}
  \caption{Secret key rate per channel use under the restricted
  non-adaptive product-measurement model, calculated from
  equation~(\ref{multicopy-key-rate}) and optimized over $m$. The curves assume entropy-preserving acceptance and the stated product-measurement attack class.}
  The parameters are fibre loss $0.2$~dB/km,
  $p_{\mathrm{dark}}=10^{-6}$, detector efficiency $\eta=25\%$, and visibility
  $V=98\%$.  Two-mode pure-loss PLOB capacity \cite{PLOB}, shown only as a standard-QKD reference benchmark; it is not a capacity bound on the multimode QCT protocol.
  \label{KeyrateCurve1}
\end{figure}

\subsection{Performance Analysis and Key Rate}
\label{sec:analysis}

Within the restricted non-adaptive product-measurement model, transmitting
$m$ copies increases Bob's detection probability while increasing Eve's
classical information by at most
\begin{equation}
\mu_m
=
\min\left\{
1,\frac{m}{d+1}
\right\}.
\end{equation}
Therefore
\begin{equation}
H(X|Z,S)
\geq
1-\mu_m.
\end{equation}
Under i.i.d. attacks in this restricted class and the entropy-preserving acceptance assumption, the corresponding asymptotic gross key-rate bound per channel use is

\begin{equation}
R_m
\geq
\bigl(1-P(\mathrm{abort})_m\bigr)
\left[
1
-
\mu_m
-
h_2(QBER_m)
\right]_+,
\label{multicopy-key-rate}
\end{equation}
where $[a]_+:=\max\{a,0\}$.

Here $P(\mathrm{abort})_m$ is the probability that Bob discards the round
because no detector clicks or because both detectors click, and $QBER_m$
is the conditional error probability among accepted rounds. Their
derivation is given in Appendix~\ref{App:MultiCopy}. This rate applies to the non-adaptive product-measurement class and the stated acceptance model. Section~\ref{sec:joint-attack} characterizes a stronger joint-measurement attack.

\noindent
Within the restricted product-measurement model, the linear-in-$d$
performance enhancement follows directly in the signal-limited regime.
Choose
\begin{equation}
m=\lfloor c(d+1)\rfloor,
\qquad
0<c<1.
\end{equation}
Then Eve's information penalty remains bounded by
$\mu_m\simeq c$, while for
$mT\eta\ll1$ the signal-detection probability increases approximately as
$mT\eta=\Theta(dT\eta)$. Consequently, whenever
$1-c-h_2(QBER_m)>0$, the achievable key rate is enhanced by a factor
$\Theta(d)$ relative to the single-copy rate before detector saturation
and dark counts become dominant. Three regimes are therefore
distinguished:
\begin{itemize}
    \item Constant-rate regime: at short distances, where the transmittance is high, $m$ can be optimized so that at least one photon consistently triggers a detector without leaking excess information to Eve. The secret key rate then remains essentially constant, of the order of one bit per channel use, and hence commensurate with the classical clock rate of the system.
    \item Single-copy regime: at intermediate distances, the key rate mirrors the single-copy case and scales linearly with the channel transmittance $T$, i.e.\ it decays exponentially with distance.
    \item Cut-off regime: at long distances, the probability of a successful signal detection drops below the dark count probability $p_{\mathrm{dark}}$. The QBER then rises sharply and the secret key rate drops abruptly to zero.
\end{itemize}

\subsection{Remaining general-security questions}
Theorem~\ref{thm:joint-attack} rules out extending the product-measurement leakage formula to arbitrary joint measurements at logarithmic copy counts. The remaining positive question is whether useful entropy bounds hold for smaller $m$, and whether the sufficient attack scale and constants are optimal. Structured square-root measurements are natural tools for this analysis \cite{Hausladen,HayashiNagaoka2003}. Further directions include coherent attacks across independently seeded rounds, pass-conditioned entropy under adversarial postselection, and quantitative implementations of the assumed time-lock encryption and effective storage behavior.

\begin{figure}[ht]
  \centering
  \includegraphics[width=\linewidth]{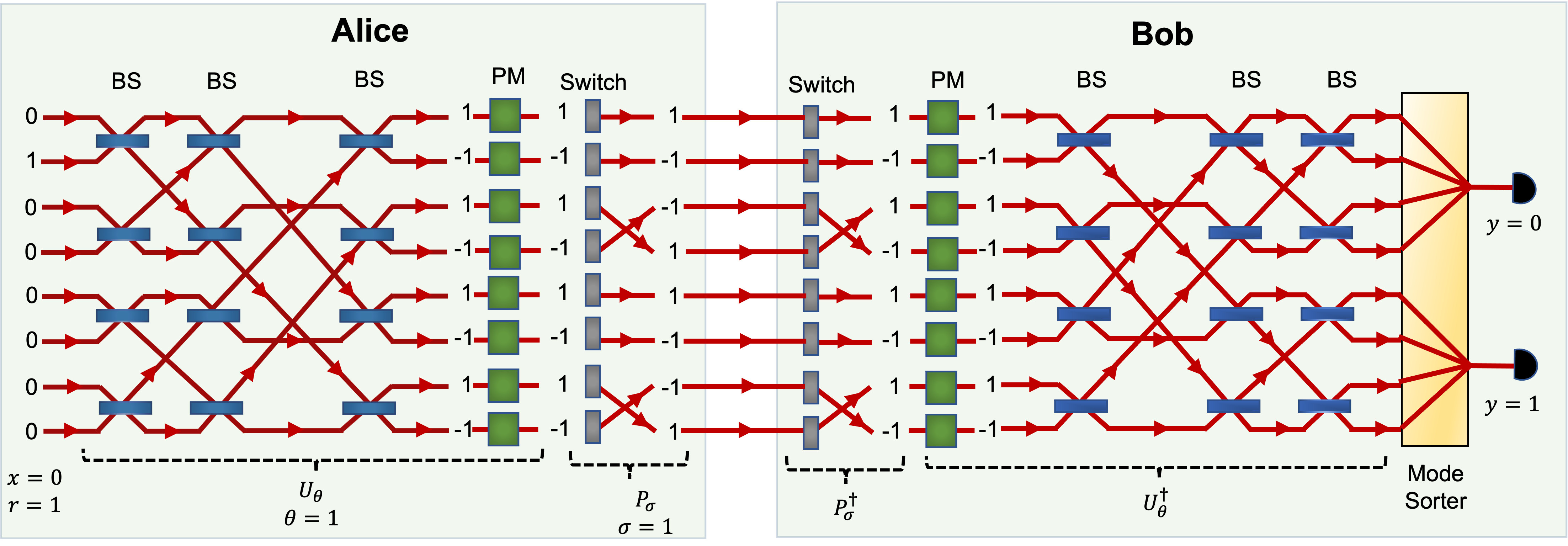}
  \caption{Conceptual spatial-mode implementation of MUB-QCT for $d=2^3$.
Alice first applies the inverse permutation $P_\sigma^\dagger$ to the
computational-basis mode and subsequently applies the MUB-preparation
unitary $U_\theta^\dagger$, thereby implementing
$V_{\theta\sigma}=U_\theta^\dagger P_\sigma^\dagger$.
Bob performs the inverse transformation
$V_{\theta\sigma}^\dagger=P_\sigma U_\theta$: he first applies
$U_\theta$ and then $P_\sigma$, before the final binary mode sorter.
The sorter partitions the $d$ computational modes into the two
rank-$d/2$ subspaces corresponding to $y=0$ and $y=1$.}
  \label{implementation}
\end{figure}

\subsection{Conceptual Physical Implementation}
\label{sec:implementation}

The experimental realization of MUB-QCT requires stable high-dimensional quantum degrees of freedom. High-dimensional encoding has been demonstrated using position-momentum \cite{PM}, temporal-spectral \cite{temporal1}, and orbital angular momentum (OAM) \cite{OAM1} modes. State-of-the-art spectral-encoding techniques indicate that dimensions $d \in [10^3, 10^8]$ are technologically accessible \cite{FLQKD2}.

We propose a conceptual implementation based on spatial modes, illustrated
in Figure~\ref{implementation}. For
$d=2^3$, Alice first prepares the computational-basis mode
$|i_{x,r}\rangle$. Alice first applies the inverse permutation
$P_\sigma^\dagger$ using the switching network and subsequently applies
the MUB-preparation transformation $U_\theta^\dagger$. The transmitted
state is therefore
\begin{equation}
|\psi_{x,r}^{\theta\sigma}\rangle
=
U_\theta^\dagger P_\sigma^\dagger|i_{x,r}\rangle,
\end{equation}
in agreement with equation~(\ref{pureencodedstate}).

Upon receiving the state, Bob, who holds the decrypted basis information $(\theta, \sigma)$, performs the inverse unitary operations. The final decoding step uses a mode sorter that groups the $d$ spatial modes into two distinct sets of $d/2$ modes, each directed to one of the two detectors. A detection event in the ``correct'' or in the ``wrong'' detector determines Bob's measurement outcome. This gives two logical outcomes while the decoding network still processes $d$ modes. The high-loss rate improvement discussed above additionally depends on the restricted attack and acceptance assumptions.

\noindent \textbf{Implementation assumptions:} The hardware model includes trusted laboratories, controlled side-channel leakage, and the specified receiver acceptance and timing behavior. With adversarially correlated losses, secrecy is quantified through an entropy bound conditioned on the monitored acceptance and timing transcript.

\section{Conclusion}
\label{conclusion}
We introduced MUB-QCT using complete MUBs and affine pair-wise independent permutations. Under the explicit time-lock encryption assumption, the single-copy measurement bound is $1/(2\sqrt{d+1})$ in trace distance and exactly $1/(d+1)$ in optimal mutual information. The effective storage bound adds at most $\delta$ to individual-attack trace distance and a continuity correction $g(\delta)$ to the entropy penalty. The asymptotic key rate is obtained for i.i.d. individual attacks, the protected timing schedule and entropy-preserving acceptance.

For repeated, independently randomized signals sharing a bit and seed, the optimal non-adaptive product-measurement leakage is $\min\{1,m/(d+1)\}$. A seed-independent joint measurement recovers the raw bit with error at most $\varepsilon$ from $O(\log d+\log(1/\varepsilon))$ signals and retains only classical information until disclosure. These results quantify the separation between product and joint measurements: the linear-in-$d$ detection enhancement belongs to the product-measurement model, while the joint attack gives a logarithmic upper bound on raw-bit recovery copy complexity. Positive joint-measurement security at smaller copy counts and optimal attack thresholds are open directions.

A promising path towards greater resistance to joint attacks is provided by the Hidden-Matching QCT (HM-QCT) protocol of Mazzoncini, Bauer, Brown and All\'eaume \cite{Mazzoncini2025hybridquantum}. Their communication-complexity construction establishes security under the QCT assumptions against arbitrary i.i.d. attacks, allowing joint processing of the signals within each channel use. Writing the number of optical modes as $d$, their result permits a photon number per channel use scaling as $O(\sqrt d/\log d)$. HM-QCT therefore points towards improving the logarithmic copy resistance exhibited here to a square-root scale, up to a logarithmic factor. This comparison motivates developing QCT encodings with stronger resistance to repeated-signal attacks, together with the corresponding implementation and key-rate analysis.

The construction is related to quantum data locking and time-limited storage \cite{QDLDV,QDLCV,CosmoLupo}, while final-key secrecy is expressed in composable trace distance. Time-lock encryption gives Bob immediate decryption and Eve delayed plaintext access; effective storage bounds the quantum information surviving until disclosure. The resulting secrecy criterion is preserved under subsequent unrestricted processing \cite{Unr13}. Extensions include coherent attacks across independently seeded rounds, finite-key pass-conditioned analysis and quantitative implementations of the time-lock encryption assumption. Two logical detection outcomes simplify the final readout, complemented by dimension-dependent mode processing.

\section*{Acknowledgments}
\label{ack}
The authors gratefully acknowledge support by the European Innovative Training Network QCALL (project 675662) funded by the Marie Sk\l odowska-Curie call H2020-MSCA-ITN-2015, and by the Quantum Technology Flagship project CiViQ, funded by the European Union's Horizon 2020 research and innovation programme under grant agreement No 820466.

\section*{Data Availability Statement}
No new data were created or analysed in this study.

\section*{Conflict of Interest}
The authors declare no competing interests.

\bibliographystyle{iopart-num}
\bibliography{biblio}

@misc{PatersonWhyQC,
      title={Quantum cryptography: a practical information security perspective}, 
      author={Kenneth G. Paterson and Fred Piper and Ruediger Schack},
      year={2009},
      eprint={quant-ph/0406147},
      archivePrefix={arXiv},
      primaryClass={quant-ph},
      url={https://arxiv.org/abs/quant-ph/0406147}, 
}

@book{Helstrom1976,
  author    = {Carl W. Helstrom},
  title     = {Quantum Detection and Estimation Theory},
  publisher = {Academic Press},
  year      = {1976}
}

@inproceedings{MUB2design,
  author    = {Andreas Klappenecker and Martin R{\"o}tteler},
  title     = {Mutually Unbiased Bases are Complex Projective 2-Designs},
  booktitle = {Proceedings of the 2005 IEEE International Symposium on Information Theory},
  pages     = {1740--1744},
  year      = {2005},
  doi       = {10.1109/ISIT.2005.1523643}
}

@article{WinterContinuity,
  author  = {Andreas Winter},
  title   = {Tight Uniform Continuity Bounds for Quantum Entropies:
             Conditional Entropy, Relative Entropy Distance and Energy Constraints},
  journal = {Communications in Mathematical Physics},
  volume  = {347},
  pages   = {291--313},
  year    = {2016},
  doi     = {10.1007/s00220-016-2609-8}
}

@article{DevetakWinter,
  author  = {Igor Devetak and Andreas Winter},
  title   = {Distillation of Secret Key and Entanglement from Quantum States},
  journal = {Proceedings of the Royal Society A: Mathematical, Physical and Engineering Sciences},
  volume  = {461},
  number  = {2053},
  pages   = {207--235},
  year    = {2005},
  doi     = {10.1098/rspa.2004.1372},
  eprint  = {quant-ph/0306078},
  archivePrefix = {arXiv}
}

@article{FullyQuantumAEP,
  author  = {Marco Tomamichel and Roger Colbeck and Renato Renner},
  title   = {A Fully Quantum Asymptotic Equipartition Property},
  journal = {IEEE Transactions on Information Theory},
  volume  = {55},
  number  = {12},
  pages   = {5840--5847},
  year    = {2009},
  doi     = {10.1109/TIT.2009.2032797},
  eprint  = {0811.1221},
  archivePrefix = {arXiv},
  primaryClass = {quant-ph}
}

@article{LeftoverHash,
  author  = {Marco Tomamichel and Christian Schaffner and Adam Smith and Renato Renner},
  title   = {Leftover Hashing Against Quantum Side Information},
  journal = {IEEE Transactions on Information Theory},
  volume  = {57},
  number  = {8},
  pages   = {5524--5535},
  year    = {2011},
  doi     = {10.1109/TIT.2011.2158473},
  eprint  = {1002.2436},
  archivePrefix = {arXiv},
  primaryClass = {quant-ph}
}

@article{TCS14,
title = {Using quantum key distribution for cryptographic purposes: A survey},
journal = {Theoretical Computer Science},
volume = {560},
pages = {62-81},
year = {2014},
note = {Theoretical Aspects of Quantum Cryptography – celebrating 30 years of BB84},
issn = {0304-3975},
doi = {https://doi.org/10.1016/j.tcs.2014.09.018},
url = {https://www.sciencedirect.com/science/article/pii/S0304397514006963},
author = {R. Alléaume and C. Branciard and J. Bouda and T. Debuisschert and M. Dianati and N. Gisin and M. Godfrey and P. Grangier and T. Länger and N. Lütkenhaus and C. Monyk and P. Painchault and M. Peev and A. Poppe and T. Pornin and J. Rarity and R. Renner and G. Ribordy and M. Riguidel and L. Salvail and A. Shields and H. Weinfurter and A. Zeilinger}
}

@inproceedings{McGrewPQC2015,
  author = {D. McGrew},
  title = {Living with postquantum cryptography},
  booktitle = {NIST Workshop on Cybersecurity in a Post-Quantum World},
  year = {2015},
  url = {http://csrc.nist.gov/groups/ST/post-quantum-2015/presentations/session4-mcgrew-david.pdf}
}

@techreport{NCSC20,
  author = {{National Cyber Security Centre}},
  title = {NCSC Whitepaper on quantum security technologies},
  institution = {NCSC},
  year = {2020},
  url = {https://www.ncsc.gov.uk/whitepaper/quantum-security-technologies}
}

@article{Advances19,
    title = {Advances in quantum cryptography},
    author = {S. Pirandola and U. L. Andersen and L. Banchi and M. Berta and D. Bunandar and R. Colbeck and D. Englund and T. Gehring and C. Lupo and C. Ottaviani and J. L. Pereira and M. Razavi and J. Shamsul Shaari and M. Tomamichel and V. C. Usenko and G. Vallone and P. Villoresi and P. Wallden},
    journal = {Adv. Opt. Photon.},
    number = {4},
    pages = {1012--1236},
    publisher = {Optica Publishing Group},
    volume = {12},
    month = {Dec},
    year = {2020},
    url = {https://opg.optica.org/aop/abstract.cfm?URI=aop-12-4-1012},
    doi = {10.1364/AOP.361502},
}

@article{SECOQCWP,
doi = {10.1088/1367-2630/11/7/075001},
url = {https://doi.org/10.1088/1367-2630/11/7/075001},
year = {2009},
month = {jul},
publisher = {},
volume = {11},
number = {7},
pages = {075001},
author = {Peev, M and Pacher, C and Alléaume, R and Barreiro, C and Bouda, J and Boxleitner, W and Debuisschert, T and Diamanti, E and Dianati, M and Dynes, J F and Fasel, S and Fossier, S and Fürst, M and Gautier, J-D and Gay, O and Gisin, N and Grangier, P and Happe, A and Hasani, Y and Hentschel, M and Hübel, H and Humer, G and Länger, T and Legré, M and Lieger, R and Lodewyck, J and Lorünser, T and Lütkenhaus, N and Marhold, A and Matyus, T and Maurhart, O and Monat, L and Nauerth, S and Page, J-B and Poppe, A and Querasser, E and Ribordy, G and Robyr, S and Salvail, L and Sharpe, A W and Shields, A J and Stucki, D and Suda, M and Tamas, C and Themel, T and Thew, R T and Thoma, Y and Treiber, A and Trinkler, P and Tualle-Brouri, R and Vannel, F and Walenta, N and Weier, H and Weinfurter, H and Wimberger, I and Yuan, Z L and Zbinden, H and Zeilinger, A},
title = {The SECOQC quantum key distribution network in Vienna},
journal = {New Journal of Physics}
}

@article{Sasaki11,
   title={Field test of quantum key distribution in the Tokyo QKD Network},
   volume={19},
   ISSN={1094-4087},
   url={http://dx.doi.org/10.1364/OE.19.010387},
   DOI={10.1364/oe.19.010387},
   number={11},
   journal={Optics Express},
   publisher={Optica Publishing Group},
   author={Sasaki, M. and Fujiwara, M. and Ishizuka, H. and Klaus, W. and Wakui, K. and Takeoka, M. and Miki, S. and Yamashita, T. and Wang, Z. and Tanaka, A. and Yoshino, K. and Nambu, Y. and Takahashi, S. and Tajima, A. and Tomita, A. and Domeki, T. and Hasegawa, T. and Sakai, Y. and Kobayashi, H. and Asai, T. and Shimizu, K. and Tokura, T. and Tsurumaru, T. and Matsui, M. and Honjo, T. and Tamaki, K. and Takesue, H. and Tokura, Y. and Dynes, J. F. and Dixon, A. R. and Sharpe, A. W. and Yuan, Z. L. and Shields, A. J. and Uchikoga, S. and Legré, M. and Robyr, S. and Trinkler, P. and Monat, L. and Page, J.-B. and Ribordy, G. and Poppe, A. and Allacher, A. and Maurhart, O. and Länger, T. and Peev, M. and Zeilinger, A.},
   year={2011},
   month=may, pages={10387} }

@article{NPJ16,
   title={Practical challenges in quantum key distribution},
   volume={2},
   ISSN={2056-6387},
   url={http://dx.doi.org/10.1038/npjqi.2016.25},
   DOI={10.1038/npjqi.2016.25},
   number={1},
   journal={npj Quantum Information},
   publisher={Springer Science and Business Media LLC},
   author={Diamanti, Eleni and Lo, Hoi-Kwong and Qi, Bing and Yuan, Zhiliang},
   year={2016},
   month=nov }

@article{QKDDeployZhang18,
   title={Large scale quantum key distribution: challenges and solutions [Invited]},
   volume={26},
   ISSN={1094-4087},
   url={http://dx.doi.org/10.1364/OE.26.024260},
   DOI={10.1364/oe.26.024260},
   number={18},
   journal={Optics Express},
   publisher={Optica Publishing Group},
   author={Zhang, Qiang and Xu, Feihu and Chen, Yu-Ao and Peng, Cheng-Zhi and Pan, Jian-Wei},
   year={2018},
   month=aug, pages={24260} }

@misc{RennerPhD,
      title={Security of Quantum Key Distribution}, 
      author={Renato Renner},
      year={2006},
      eprint={quant-ph/0512258},
      archivePrefix={arXiv},
      primaryClass={quant-ph},
      url={https://arxiv.org/abs/quant-ph/0512258}, 
}

@article{ScaraniRMP09,
   title={The security of practical quantum key distribution},
   volume={81},
   ISSN={1539-0756},
   url={http://dx.doi.org/10.1103/RevModPhys.81.1301},
   DOI={10.1103/revmodphys.81.1301},
   number={3},
   journal={Reviews of Modern Physics},
   publisher={American Physical Society (APS)},
   author={Scarani, Valerio and Bechmann-Pasquinucci, Helle and Cerf, Nicolas J. and Dušek, Miloslav and Lütkenhaus, Norbert and Peev, Momtchil},
   year={2009},
   month=sep, pages={1301–1350} }

@article{Toma17,
   title={A largely self-contained and complete security proof for quantum key distribution},
   volume={1},
   ISSN={2521-327X},
   url={http://dx.doi.org/10.22331/q-2017-07-14-14},
   DOI={10.22331/q-2017-07-14-14},
   journal={Quantum},
   publisher={Verein zur Forderung des Open Access Publizierens in den Quantenwissenschaften},
   author={Tomamichel, Marco and Leverrier, Anthony},
   year={2017},
   month=jul, pages={14} }

@article{ETSIWP,
  title={Implementation Security of Quantum Cryptography},
  author={Lucamarini, Marco and others},
  journal={ETSI White Paper},
  number={27},
  year={2018},
  publisher={ETSI},
  url={https://www.etsi.org/images/files/ETSIWhitePapers/etsi_wp27_qkd_imp_sec_FINAL.pdf}
}

@article{Xu19,
   title={Secure quantum key distribution with realistic devices},
   volume={92},
   ISSN={1539-0756},
   url={http://dx.doi.org/10.1103/RevModPhys.92.025002},
   DOI={10.1103/revmodphys.92.025002},
   number={2},
   journal={Reviews of Modern Physics},
   publisher={American Physical Society (APS)},
   author={Xu, Feihu and Ma, Xiongfeng and Zhang, Qiang and Lo, Hoi-Kwong and Pan, Jian-Wei},
   year={2020},
   month=may }

@misc{BlackPaper,
      title={The black paper of quantum cryptography: real implementation problems}, 
      author={Valerio Scarani and Christian Kurtsiefer},
      year={2012},
      eprint={0906.4547},
      archivePrefix={arXiv},
      primaryClass={quant-ph},
      url={https://arxiv.org/abs/0906.4547}, 
}

@article{PLOB,
   title={Fundamental limits of repeaterless quantum communications},
   volume={8},
   ISSN={2041-1723},
   url={http://dx.doi.org/10.1038/ncomms15043},
   DOI={10.1038/ncomms15043},
   number={1},
   journal={Nature Communications},
   publisher={Springer Science and Business Media LLC},
   author={Pirandola, Stefano and Laurenza, Riccardo and Ottaviani, Carlo and Banchi, Leonardo},
   year={2017},
   month=apr }

@article{Unr13,
  author  = {Unruh, Dominique},
  title   = {Everlasting Multi-party Computation},
  journal = {Journal of Cryptology},
  year    = {2018},
  month   = {10},
  volume  = {31},
  number  = {4},
  pages   = {965--1011},
  doi     = {10.1007/s00145-018-9278-z},
  url     = {https://doi.org/10.1007/s00145-018-9278-z},
  issn    = {1432-1378}
}

@phdthesis{AllHDR21,
  TITLE = {{Quantum cryptography and its application frontiers}},
  AUTHOR = {Alleaume, Romain},
  URL = {https://theses.hal.science/tel-04255325},
  SCHOOL = {{Sorbonne Universit{\'e} (France)}},
  YEAR = {2021},
  MONTH = Sep,
  TYPE = {Accreditation to supervise research},
  HAL_ID = {tel-04255325},
  HAL_VERSION = {v2},
}

@article{GisinRMP02,
   title={Quantum cryptography},
   volume={74},
   ISSN={1539-0756},
   url={http://dx.doi.org/10.1103/RevModPhys.74.145},
   DOI={10.1103/revmodphys.74.145},
   number={1},
   journal={Reviews of Modern Physics},
   publisher={American Physical Society (APS)},
   author={Gisin, Nicolas and Ribordy, Grégoire and Tittel, Wolfgang and Zbinden, Hugo},
   year={2002},
   month=mar, pages={145–195} }

@techreport{ETSIQKDPP,
  author = {ETSI},
  title = {Quantum Key Distribution (QKD); Common Criteria Protection Profile - Pair of Prepare and Measure Quantum Key Distribution Modules},
  institution = {European Telecommunications Standards Institute (ETSI)},
  year = {2024},
  type = {Group Specification},
  number = {ETSI GS QKD 016 V2.1.1},
  url = {https://www.etsi.org/deliver/etsi_gs/QKD/001_099/016/02.01.01_60/gs_qkd016v020101p.pdf}
}

@article{TRE,
author = {Unruh, Dominique},
title = {Revocable Quantum Timed-Release Encryption},
year = {2015},
issue_date = {December 2015},
publisher = {Association for Computing Machinery},
address = {New York, NY, USA},
volume = {62},
number = {6},
issn = {0004-5411},
url = {https://doi.org/10.1145/2817206},
doi = {10.1145/2817206},
journal = {J. ACM},
month = dec,
articleno = {49},
numpages = {76}
}

@misc{Unr06,
author = "Unruh, Dominique",
title = "Protocol Composition and Complexity",
year = 2007,
howpublished = "Excellent Computer Science Dissertations 2006",
publisher = "Gesellschaft für Informatik",
address = "Bonn",
pissn = "1617-5468",
isbn = "978-3-88579-411-0",
pages = "189--198",
}

@article{NoisyStorage,
   title={Cryptography from Noisy Storage},
   volume={100},
   ISSN={1079-7114},
   url={http://dx.doi.org/10.1103/PhysRevLett.100.220502},
   DOI={10.1103/physrevlett.100.220502},
   number={22},
   journal={Physical Review Letters},
   publisher={American Physical Society (APS)},
   author={Wehner, Stephanie and Schaffner, Christian and Terhal, Barbara M.},
   year={2008},
   month=jun }

@article{QDLDV,
   title={Quantum-Locked Key Distribution at Nearly the Classical Capacity Rate},
   volume={113},
   ISSN={1079-7114},
   url={http://dx.doi.org/10.1103/PhysRevLett.113.160502},
   DOI={10.1103/physrevlett.113.160502},
   number={16},
   journal={Physical Review Letters},
   publisher={American Physical Society (APS)},
   author={Lupo, Cosmo and Lloyd, Seth},
   year={2014},
   month=oct }

@article{QDLCV,
   title={Continuous-variable quantum enigma machines for long-distance key distribution},
   volume={92},
   ISSN={1094-1622},
   url={http://dx.doi.org/10.1103/PhysRevA.92.062312},
   DOI={10.1103/physreva.92.062312},
   number={6},
   journal={Physical Review A},
   publisher={American Physical Society (APS)},
   author={Lupo, Cosmo and Lloyd, Seth},
   year={2015},
   month=dec }

@Article{CosmoLupo,
AUTHOR = {Lupo, Cosmo},
TITLE = {Quantum Data Locking for Secure Communication against an Eavesdropper with Time-Limited Storage},
JOURNAL = {Entropy},
VOLUME = {17},
YEAR = {2015},
NUMBER = {5},
PAGES = {3194--3204},
URL = {https://www.mdpi.com/1099-4300/17/5/3194},
ISSN = {1099-4300},
DOI = {10.3390/e17053194}
}

@article{FLQKD2,
   title={Floodlight quantum key distribution: Demonstrating a framework for high-rate secure communication},
   volume={95},
   ISSN={2469-9934},
   url={http://dx.doi.org/10.1103/PhysRevA.95.012332},
   DOI={10.1103/physreva.95.012332},
   number={1},
   journal={Physical Review A},
   publisher={American Physical Society (APS)},
   author={Zhang, Zheshen and Zhuang, Quntao and Wong, Franco N. C. and Shapiro, Jeffrey H.},
   year={2017},
   month=jan }

@techreport{topsecret,
  author = {Elaine B. Barker},
  title = {{Recommendation for Key Management, Part 1: General (Revision 5)}},
  institution = {National Institute of Standards and Technology ({NIST})},
  type = {Special Publication ({SP})},
  number = {800-57 Part 1 Revision 5},
  year = {2020},
  month = {may},
  doi = {10.6028/NIST.SP.800-57pt1r5},
  url = {https://csrc.nist.gov/pubs/sp/800/57/pt1/r5/final}
}

@article{TIon,
   title={Single-qubit quantum memory exceeding ten-minute coherence time},
   volume={11},
   ISSN={1749-4893},
   url={http://dx.doi.org/10.1038/s41566-017-0007-1},
   DOI={10.1038/s41566-017-0007-1},
   number={10},
   journal={Nature Photonics},
   publisher={Springer Science and Business Media LLC},
   author={Wang, Ye and Um, Mark and Zhang, Junhua and An, Shuoming and Lyu, Ming and Zhang, Jing-Ning and Duan, L.-M. and Yum, Dahyun and Kim, Kihwan},
   year={2017},
   month=sep, pages={646–650} }

@article{Tion1,
   title={Remote Entanglement between a Single Atom and a Bose-Einstein Condensate},
   volume={106},
   ISSN={1079-7114},
   url={http://dx.doi.org/10.1103/PhysRevLett.106.210503},
   DOI={10.1103/physrevlett.106.210503},
   number={21},
   journal={Physical Review Letters},
   publisher={American Physical Society (APS)},
   author={Lettner, M. and Mücke, M. and Riedl, S. and Vo, C. and Hahn, C. and Baur, S. and Bochmann, J. and Ritter, S. and Dürr, S. and Rempe, G.},
   year={2011},
   month=may }

@article{NV,
   title={Coherent control of the silicon-vacancy spin in diamond},
   volume={8},
   ISSN={2041-1723},
   url={http://dx.doi.org/10.1038/ncomms15579},
   DOI={10.1038/ncomms15579},
   number={1},
   journal={Nature Communications},
   publisher={Springer Science and Business Media LLC},
   author={Pingault, Benjamin and Jarausch, David-Dominik and Hepp, Christian and Klintberg, Lina and Becker, Jonas N. and Markham, Matthew and Becher, Christoph and Atatüre, Mete},
   year={2017},
   month=may }

@article{NVC,
   title={Broadband noise-free optical quantum memory with neutral nitrogen-vacancy centers in diamond},
   volume={91},
   ISSN={1550-235X},
   url={http://dx.doi.org/10.1103/PhysRevB.91.205108},
   DOI={10.1103/physrevb.91.205108},
   number={20},
   journal={Physical Review B},
   publisher={American Physical Society (APS)},
   author={Poem, E. and Weinzetl, C. and Klatzow, J. and Kaczmarek, K. T. and Munns, J. H. D. and Champion, T. F. M. and Saunders, D. J. and Nunn, J. and Walmsley, I. A.},
   year={2015},
   month=may }

@article{NVC1,
  author  = {Awschalom, David D. and Hanson, Ronald and Wrachtrup, J{\"o}rg and Zhou, Brian B.},
  title   = {Quantum technologies with optically interfaced solid-state spins},
  journal = {Nature Photonics},
  year    = {2018},
  month   = {9},
  volume  = {12},
  number  = {9},
  pages   = {516--527},
  doi     = {10.1038/s41566-018-0232-2},
  url     = {https://doi.org/10.1038/s41566-018-0232-2},
  issn    = {1749-4893}
}

@article{QED,
author = {M. H. Devoret  and R. J. Schoelkopf },
title = {Superconducting Circuits for Quantum Information: An Outlook},
journal = {Science},
volume = {339},
number = {6124},
pages = {1169-1174},
year = {2013},
doi = {10.1126/science.1231930},
URL = {https://www.science.org/doi/abs/10.1126/science.1231930},
eprint = {https://www.science.org/doi/pdf/10.1126/science.1231930}}

@article{QED1,
  title = {Heralded Storage of a Photonic Quantum Bit in a Single Atom},
  author = {Kalb, Norbert and Reiserer, Andreas and Ritter, Stephan and Rempe, Gerhard},
  journal = {Phys. Rev. Lett.},
  volume = {114},
  issue = {22},
  pages = {220501},
  numpages = {5},
  year = {2015},
  month = {Jun},
  publisher = {American Physical Society},
  doi = {10.1103/PhysRevLett.114.220501},
  url = {https://link.aps.org/doi/10.1103/PhysRevLett.114.220501}
}

@article{Raman1,
   title={Towards high-speed optical quantum memories},
   volume={4},
   ISSN={1749-4893},
   url={http://dx.doi.org/10.1038/nphoton.2010.30},
   DOI={10.1038/nphoton.2010.30},
   number={4},
   journal={Nature Photonics},
   publisher={Springer Science and Business Media LLC},
   author={Reim, K. F. and Nunn, J. and Lorenz, V. O. and Sussman, B. J. and Lee, K. C. and Langford, N. K. and Jaksch, D. and Walmsley, I. A.},
   year={2010},
   month=mar, pages={218–221} }

@article{Raman2,
  title = {Single-Photon-Level Quantum Memory at Room Temperature},
  author = {Reim, K. F. and Michelberger, P. and Lee, K. C. and Nunn, J. and Langford, N. K. and Walmsley, I. A.},
  journal = {Phys. Rev. Lett.},
  volume = {107},
  issue = {5},
  pages = {053603},
  numpages = {4},
  year = {2011},
  month = {Jul},
  publisher = {American Physical Society},
  doi = {10.1103/PhysRevLett.107.053603},
  url = {https://link.aps.org/doi/10.1103/PhysRevLett.107.053603}
}

@article{Raman3,
   title={High-performance Raman memory with spatio-temporal reversal},
   volume={26},
   ISSN={1094-4087},
   url={http://dx.doi.org/10.1364/OE.26.012424},
   DOI={10.1364/oe.26.012424},
   number={10},
   journal={Optics Express},
   publisher={Optica Publishing Group},
   author={Vernaz-Gris, Pierre and Tranter, Aaron D. and Everett, Jesse L. and Leung, Anthony C. and Paul, Karun V. and Campbell, Geoff T. and Lam, Ping Koy and Buchler, Ben C.},
   year={2018},
   month=may, pages={12424} }

@article{AFC,
  title = {Coherent Spin Control at the Quantum Level in an Ensemble-Based Optical Memory},
  author = {Jobez, Pierre and Laplane, Cyril and Timoney, Nuala and Gisin, Nicolas and Ferrier, Alban and Goldner, Philippe and Afzelius, Mikael},
  journal = {Phys. Rev. Lett.},
  volume = {114},
  issue = {23},
  pages = {230502},
  numpages = {5},
  year = {2015},
  month = {Jun},
  publisher = {American Physical Society},
  doi = {10.1103/PhysRevLett.114.230502},
  url = {https://link.aps.org/doi/10.1103/PhysRevLett.114.230502}
}

@article{AFC1,
   title={Simultaneous coherence enhancement of optical and microwave transitions in solid-state electronic spins},
   volume={17},
   ISSN={1476-4660},
   url={http://dx.doi.org/10.1038/s41563-018-0138-x},
   DOI={10.1038/s41563-018-0138-x},
   number={8},
   journal={Nature Materials},
   publisher={Springer Science and Business Media LLC},
   author={Ortu, Antonio and Tiranov, Alexey and Welinski, Sacha and Fröwis, Florian and Gisin, Nicolas and Ferrier, Alban and Goldner, Philippe and Afzelius, Mikael},
   year={2018},
   month=jul, pages={671–675} }

@article{AFC2,
   title={Optical storage for 0.53 s in a solid-state atomic frequency comb memory using dynamical decoupling},
   volume={22},
   ISSN={1367-2630},
   url={http://dx.doi.org/10.1088/1367-2630/ab8aac},
   DOI={10.1088/1367-2630/ab8aac},
   number={6},
   journal={New Journal of Physics},
   publisher={IOP Publishing},
   author={Holzäpfel, Adrian and Etesse, Jean and Kaczmarek, Krzysztof T and Tiranov, Alexey and Gisin, Nicolas and Afzelius, Mikael},
   year={2020},
   month=jun, pages={063009} }

@article{Hausladen,
author = {Paul Hausladen and William K. Wootters},
title = {A ‘Pretty Good’ Measurement for Distinguishing Quantum States},
journal = {Journal of Modern Optics},
volume = {41},
number = {12},
pages = {2385--2390},
year = {1994},
publisher = {Taylor \& Francis},
doi = {10.1080/09500349414552221},


URL = { 
    
        https://doi.org/10.1080/09500349414552221
    
    

},
eprint = { 
    
        https://doi.org/10.1080/09500349414552221
    
    

}

}

@article{Cavity,
   title={Time-reversed and coherently enhanced memory: A single-mode quantum atom-optic memory without a cavity},
   volume={98},
   ISSN={2469-9934},
   url={http://dx.doi.org/10.1103/PhysRevA.98.063846},
   DOI={10.1103/physreva.98.063846},
   number={6},
   journal={Physical Review A},
   publisher={American Physical Society (APS)},
   author={Everett, J. L. and Vernaz-Gris, P. and Campbell, G. T. and Tranter, A. D. and Paul, K. V. and Leung, A. C. and Lam, P. K. and Buchler, B. C.},
   year={2018},
   month=dec }

@article{EIT,
  title = {Coherent Optical Memory with High Storage Efficiency and Large Fractional Delay},
  author = {Chen, Yi-Hsin and Lee, Meng-Jung and Wang, I-Chung and Du, Shengwang and Chen, Yong-Fan and Chen, Ying-Cheng and Yu, Ite A.},
  journal = {Phys. Rev. Lett.},
  volume = {110},
  issue = {8},
  pages = {083601},
  numpages = {5},
  year = {2013},
  month = {Feb},
  publisher = {American Physical Society},
  doi = {10.1103/PhysRevLett.110.083601},
  url = {https://link.aps.org/doi/10.1103/PhysRevLett.110.083601}
}

@article{EIT1,
  title = {Broadband coherent optical memory based on electromagnetically induced transparency},
  author = {Wei, Yan-Cheng and Wu, Bo-Han and Hsiao, Ya-Fen and Tsai, Pin-Ju and Chen, Ying-Cheng},
  journal = {Phys. Rev. A},
  volume = {102},
  issue = {6},
  pages = {063720},
  numpages = {11},
  year = {2020},
  month = {Dec},
  publisher = {American Physical Society},
  doi = {10.1103/PhysRevA.102.063720},
  url = {https://link.aps.org/doi/10.1103/PhysRevA.102.063720}
}

@article{GEM,
author = {Y.-W. Cho and G. T. Campbell and J. L. Everett and J. Bernu and D. B. Higginbottom and M. T. Cao and J. Geng and N. P. Robins and P. K. Lam and B. C. Buchler},
journal = {Optica},
number = {1},
pages = {100--107},
publisher = {Optica Publishing Group},
title = {Highly efficient optical quantum memory with long coherence time in cold atoms},
volume = {3},
month = {Jan},
year = {2016},
url = {https://opg.optica.org/optica/abstract.cfm?URI=optica-3-1-100},
doi = {10.1364/OPTICA.3.000100},
}

@article{DLCZ,
  title = {Homodyne Tomography of a Single Photon Retrieved on Demand from a Cavity-Enhanced Cold Atom Memory},
  author = {Bimbard, Erwan and Boddeda, Rajiv and Vitrant, Nicolas and Grankin, Andrey and Parigi, Valentina and Stanojevic, Jovica and Ourjoumtsev, Alexei and Grangier, Philippe},
  journal = {Phys. Rev. Lett.},
  volume = {112},
  issue = {3},
  pages = {033601},
  numpages = {5},
  year = {2014},
  month = {Jan},
  publisher = {American Physical Society},
  doi = {10.1103/PhysRevLett.112.033601},
  url = {https://link.aps.org/doi/10.1103/PhysRevLett.112.033601}
}

@article{EC,
title = {Universal classes of hash functions},
journal = {Journal of Computer and System Sciences},
volume = {18},
number = {2},
pages = {143-154},
year = {1979},
issn = {0022-0000},
doi = {https://doi.org/10.1016/0022-0000(79)90044-8},
url = {https://www.sciencedirect.com/science/article/pii/0022000079900448},
author = {J.Lawrence Carter and Mark N. Wegman}
}

@ARTICLE{PA,
  author={Bennett, C.H. and Brassard, G. and Crepeau, C. and Maurer, U.M.},
  journal={IEEE Transactions on Information Theory}, 
  title={Generalized privacy amplification}, 
  year={1995},
  volume={41},
  number={6},
  pages={1915-1923},
  doi={10.1109/18.476316}}

@misc{Renner,
      title={Cryptographic security of quantum key distribution}, 
      author={Christopher Portmann and Renato Renner},
      year={2014},
      eprint={1409.3525},
      archivePrefix={arXiv},
      primaryClass={quant-ph},
      url={https://arxiv.org/abs/1409.3525}, 
}

@article{QCE,
   title={Quantum to Classical Randomness Extractors},
   volume={60},
   ISSN={1557-9654},
   url={http://dx.doi.org/10.1109/TIT.2013.2291780},
   DOI={10.1109/tit.2013.2291780},
   number={2},
   journal={IEEE Transactions on Information Theory},
   publisher={Institute of Electrical and Electronics Engineers (IEEE)},
   author={Berta, Mario and Fawzi, Omar and Wehner, Stephanie},
   year={2014},
   month=feb, pages={1168–1192} }

@article{HD1,
   title={Efficient decoy-state quantum key distribution with quantified security},
   volume={21},
   ISSN={1094-4087},
   url={http://dx.doi.org/10.1364/OE.21.024550},
   DOI={10.1364/oe.21.024550},
   number={21},
   journal={Optics Express},
   publisher={Optica Publishing Group},
   author={Lucamarini, M. and Patel, K. A. and Dynes, J. F. and Fröhlich, B. and Sharpe, A. W. and Dixon, A. R. and Yuan, Z. L. and Penty, R. V. and Shields, A. J.},
   year={2013},
   month=oct, pages={24550} }

@article{HD2,
doi = {10.1088/1367-2630/17/2/022002},
url = {https://doi.org/10.1088/1367-2630/17/2/022002},
year = {2015},
month = {feb},
publisher = {IOP Publishing},
volume = {17},
number = {2},
pages = {022002},
author = {Zhong, Tian and Zhou, Hongchao and Horansky, Robert D and Lee, Catherine and Verma, Varun B and Lita, Adriana E and Restelli, Alessandro and Bienfang, Joshua C and Mirin, Richard P and Gerrits, Thomas and Nam, Sae Woo and Marsili, Francesco and Shaw, Matthew D and Zhang, Zheshen and Wang, Ligong and Englund, Dirk and Wornell, Gregory W and Shapiro, Jeffrey H and Wong, Franco N C},
title = {Photon-efficient quantum key distribution using time–energy entanglement with high-dimensional encoding},
journal = {New Journal of Physics}
}

@misc{HD3,
      title={High-rate field demonstration of large-alphabet quantum key distribution}, 
      author={Catherine Lee and Darius Bunandar and Zheshen Zhang and Gregory R. Steinbrecher and P. Ben Dixon and Franco N. C. Wong and Jeffrey H. Shapiro and Scott A. Hamilton and Dirk Englund},
      year={2016},
      eprint={1611.01139},
      archivePrefix={arXiv},
      primaryClass={quant-ph},
      url={https://arxiv.org/abs/1611.01139}, 
}

@article{HD4,
   title={Efficient High-Dimensional Quantum Key Distribution with Hybrid Encoding},
   volume={21},
   ISSN={1099-4300},
   url={http://dx.doi.org/10.3390/e21010080},
   DOI={10.3390/e21010080},
   number={1},
   journal={Entropy},
   publisher={MDPI AG},
   author={Jo, Yonggi and Park, Hee Su and Lee, Seung-Woo and Son, Wonmin},
   year={2019},
   month=jan, pages={80} }

@article{PM,
  title = {Secure Quantum Key Distribution using Continuous Variables of Single Photons},
  author = {Zhang, Lijian and Silberhorn, Christine and Walmsley, Ian A.},
  journal = {Phys. Rev. Lett.},
  volume = {100},
  issue = {11},
  pages = {110504},
  numpages = {4},
  year = {2008},
  month = {Mar},
  publisher = {American Physical Society},
  doi = {10.1103/PhysRevLett.100.110504},
  url = {https://link.aps.org/doi/10.1103/PhysRevLett.100.110504}
}

@article{temporal1,
  title = {Quantum Cryptography Using Entangled Photons in Energy-Time Bell States},
  author = {Tittel, W. and Brendel, J. and Zbinden, H. and Gisin, N.},
  journal = {Phys. Rev. Lett.},
  volume = {84},
  issue = {20},
  pages = {4737--4740},
  numpages = {0},
  year = {2000},
  month = {May},
  publisher = {American Physical Society},
  doi = {10.1103/PhysRevLett.84.4737},
  url = {https://link.aps.org/doi/10.1103/PhysRevLett.84.4737}
}

@article{OAM1,
author = {Alicia Sit and Fr\'{e}d\'{e}ric Bouchard and Robert Fickler and J\'{e}r\'{e}mie Gagnon-Bischoff and Hugo Larocque and Khabat Heshami and Dominique Elser and Christian Peuntinger and Kevin G\"{u}nthner and Bettina Heim and Christoph Marquardt and Gerd Leuchs and Robert W. Boyd and Ebrahim Karimi},
journal = {Optica},
number = {9},
pages = {1006--1010},
publisher = {Optica Publishing Group},
title = {High-dimensional intracity quantum cryptography with structured photons},
volume = {4},
month = {Sep},
year = {2017},
url = {https://opg.optica.org/optica/abstract.cfm?URI=optica-4-9-1006},
doi = {10.1364/OPTICA.4.001006},
}

@misc{fullset,
      title={A new proof for the existence of mutually unbiased bases}, 
      author={Somshubhro Bandyopadhyay and P. Oscar Boykin and Vwani Roychowdhury and Farrokh Vatan},
      year={2001},
      eprint={quant-ph/0103162},
      archivePrefix={arXiv},
      primaryClass={quant-ph},
      url={https://arxiv.org/abs/quant-ph/0103162}, 
}

@article{Prime,
   title={ON MUTUALLY UNBIASED BASES},
   volume={08},
   ISSN={1793-6918},
   url={http://dx.doi.org/10.1142/S0219749910006502},
   DOI={10.1142/s0219749910006502},
   number={04},
   journal={International Journal of Quantum Information},
   publisher={World Scientific Pub Co Pte Lt},
   author={DURT, THOMAS and ENGLERT, BERTHOLD-GEORG and BENGTSSON, INGEMAR and ŻYCZKOWSKI, KAROL},
   year={2010},
   month=jun, pages={535–640} }

@misc{square,
      title={New Construction of Mutually Unbiased Bases in Square Dimensions}, 
      author={Pawel Wocjan and Thomas Beth},
      year={2004},
      eprint={quant-ph/0407081},
      archivePrefix={arXiv},
      primaryClass={quant-ph},
      url={https://arxiv.org/abs/quant-ph/0407081}, 
}

@article{HayashiNagaoka2003,
 author={Masahito Hayashi and Hiroshi Nagaoka},
 title={General formulas for capacity of classical-quantum channels},
 journal={IEEE Transactions on Information Theory},
 volume={49}, number={7}, pages={1753--1768}, year={2003},
 doi={10.1109/TIT.2003.813556},
 eprint={quant-ph/0206186}, archivePrefix={arXiv}
}

@article{Mazzoncini2025hybridquantum,
  author = {Mazzoncini, Francesco and Bauer, Balthazar and Brown, Peter and All{\'e}aume, Romain},
  title = {Hybrid {Quantum} Cryptography from {Communication} Complexity},
  journal = {Quantum},
  volume = {9},
  pages = {1862},
  year = {2025},
  doi = {10.22331/q-2025-09-24-1862},
  url = {https://doi.org/10.22331/q-2025-09-24-1862}
}
    
\appendix

\newpage

\section{Cryptographic Definitions for the QCT Model}
\label{CryptoDefinitions}
We describe two computational settings for time-lock encryption: symmetric encryption with bounded pre-expiry computational resources, and trapdoor time-release encryption with a sequential-time assumption. In both settings, the ciphertext is assumed to have the delayed-access security property of Definition~\ref{time-lock}. The definitions below describe the computational mechanism supporting the protected interval; the exact simulation property states the encryption hypothesis used by our proofs.

\begin{Definition}[Negligible function] 
A function $f$ is considered to be \textit{negligible} if for every polynomial $p(\cdot)$ there exists an $N$ such that for all integers $n>N$ it holds that $f(n)<\frac{1}{p(n)}$. 
\end{Definition}

\begin{Definition}[Standard computationally secure encryption]
\label{ComSec_Standard}
Let $\lambda$ be the security parameter and let
$k\leftarrow\operatorname{Gen}(1^\lambda)$. For every quantum polynomial-time adversary $A$ in the specified security experiment
 and every pair of equal-length messages $m_0,m_1$, define
\begin{equation}
\operatorname{Adv}^{\mathrm{ind}}_{\Pi,A}(\lambda)
:=
\left|
\Pr[A(\operatorname{Enc}_{k}(m_0))=1]
-
\Pr[A(\operatorname{Enc}_{k}(m_1))=1]
\right|.
\end{equation}
The scheme is computationally indistinguishable if
\begin{equation}
\operatorname{Adv}^{\mathrm{ind}}_{\Pi,A}(\lambda)
\leq
\operatorname{negl}(\lambda)
\end{equation}
for every such adversary.
\end{Definition}

\paragraph{Symmetric time-lock encryption assumption.}
For the symmetric setting, we assume a secret encryption key shared by Alice and Bob and a fixed bound on Eve's computational and parallel resources during the protected interval. Under that bound we assume quantum computational confidentiality for time $t_{comp}$ and the delayed-access security property of Definition~\ref{time-lock}. Bob decrypts immediately with the shared key. After expiry Eve receives the seed and may process her retained information without computational restriction. AES is a possible engineering choice for this setting; the protocol-level security statement uses the explicitly stated time-lock encryption assumption.

\paragraph{Sequential-time setting.}
For trapdoor time-release encryption we assume a sequential-work lower bound against the specified pre-expiry quantum adversary class \cite{TRE}, together with the delayed-access security property of Definition~\ref{time-lock}. A model of quantum operation rates and parallel resources converts the sequential-work parameter into the protected duration $t_{comp}$.

\begin{Definition}[Trapdoor time-lock encryption assumption]
\label{ComSec_TRE}
We assume a trapdoor time-lock encryption scheme with the delayed-access security property of Definition~\ref{time-lock} and the following computational decryption asymmetry. An authorized receiver holding a trapdoor
$\mathsf{td}_B$ can recover the plaintext immediately,
\begin{equation}
\operatorname{Dec}_{\mathsf{td}_B}(c)=m,
\end{equation}
whereas an adversary without the trapdoor must perform an inherently
sequential computation whose elapsed real-world time is at least
$t_{comp}$. In particular, for every adversary $A$ running for
time $t<t_{comp}$,
\begin{equation}
\operatorname{Adv}^{\mathrm{ind}}_{A}(t)
\leq
\operatorname{negl}(\lambda).
\end{equation}
\end{Definition}

The trapdoor supplies Bob with immediate authorized decryption while the sequential-work assumption sets Eve's protected interval. Symmetric time-lock encryption and trapdoor time-lock encryption thus share the same delayed-access security assumption in the main proof, with different computational mechanisms for authorized and unauthorized access.

\paragraph{Classical transcript and hardware assumption.}
The delayed-access property of Definition~\ref{time-lock} covers Eve's ciphertexts and encryption-device side information. At expiry, her complete encryption-related view has the representation in equation~(\ref{eq:time-lock-view}). The quantum signal randomizers remain local to Alice. Public parameter estimation and reconciliation occur according to Protocol~1 and their leakage is accounted for in Lemma~\ref{lem:privacy-amplification}.



\section{Quantum-to-Classical Randomness Extractor}
\label{QCExtractor}

\begin{Definition}[QC-extractor \cite{QCE}]
Let $A=A_{1}A_{2}$ be a bipartite quantum system with total dimension $|A|$, and let $q=\log_2 |A|$. For a min-entropy bound $k\in [-q,q]$ and an error parameter $\varepsilon\in [0,1]$, a ($k,\varepsilon$)-QC-extractor is defined by a set of unitary transformations $\{ U_{1},\ldots, U_{L}\}$ acting on $A$. For any joint state $\rho_{AE}$ satisfying $H_{\min}(A|E)\geq k$, the extractor guarantees that:
\begin{equation}
\frac{1}{L}\sum_{j=1}^{L}\Big|\Big|\hat{\Pi}_{A\rightarrow X_{1}}^{j}(\rho_{AE}) - \frac{\mathbb{I}_{A_{1}}}{|A_{1}|} \otimes \rho_{E} \Big|\Big|_1 \leq \varepsilon.
\label{QCE}
\end{equation}
\label{QCext}
\end{Definition}

\noindent The measurement map $\hat{\Pi}_{A\rightarrow X_{1}}^{j}$ is formally defined as:
\begin{equation}
\hat{\Pi}_{A\rightarrow X_{1}}^{j}(\rho_{AE}) := \mathcal{I}_{A_{1}\rightarrow X_{1}} \Big( \mathcal{T}_{A\rightarrow A_{1}} \Big( U_{j}\rho_{AE}U^{\dagger}_{j} \Big)\Big)
\label{QCMeasurement}
\end{equation}
where $\mathcal{I}_{A_{1}\rightarrow X_{1}}$ is the canonical identity map embedding the measurement result into a classical register (i.e., $\mathcal{I}_{A_{1}\rightarrow X_{1}} (|i\rangle\langle j|_{A_{1}})= |i\rangle\langle j|_{X_{1}}$), and $\mathcal{T}_{A\rightarrow A_{1}}$ is the completely dephasing partial measurement map:
\begin{equation}
\mathcal{T}_{A\rightarrow A_{1}}(\cdot) = \sum_{a_{1}\in A_{1}, a_{2}\in A_{2}} \langle a_{1}a_{2}| (\cdot)|a_{1}a_{2}\rangle |a_{1}\rangle\langle a_{1}|
\label{measurementmap}
\end{equation}
with $\{|a_{1}\rangle\}$ and $\{|a_{2}\rangle\}$ denoting the standard orthonormal bases of subsystems $A_{1}$ and $A_{2}$, respectively.


\noindent
For the prime-power dimensions considered here, identify the computational
labels with the finite field $\mathrm{GF}(d)$ and use the standard affine
pair-wise independent permutation family \cite{QCE},
\begin{equation}
\mathcal{P}_{\mathrm{aff}}
=
\left\{
\pi_{a,b}:i\mapsto ai+b
\;:\;
a\in\mathrm{GF}(d)^{\times},
\;
b\in\mathrm{GF}(d)
\right\}.
\end{equation}
For any distinct $i_1,i_2$ and distinct $j_1,j_2$, there is exactly one
pair $(a,b)$ satisfying
$\pi_{a,b}(i_1)=j_1$ and $\pi_{a,b}(i_2)=j_2$. Hence this family is
pair-wise independent and has
\begin{equation}
|\mathcal{P}_{\mathrm{aff}}|
=
d(d-1).
\end{equation}
We use $\mathcal{P}:=\mathcal{P}_{\mathrm{aff}}$ throughout. The complete
seed $(\theta,\sigma)$ therefore requires
\begin{equation}
\log_2(d+1)+\log_2 d+\log_2(d-1)
=
3\log_2 d+O(1/d)
\end{equation}
bits, rather than a quantity linear in $d$.

The following theorem is Theorem III.8 of
Berta, Fawzi and Wehner \cite{QCE}, specialized to zero smoothing.
\begin{theorem}[QC-extractor based on a full set of MUBs]
\label{QCMUB}
Let $A=A_{1}A_{2}$ be a quantum system of prime-power dimension $|A|$,
let $q=\log_2|A|$, and let $\mathcal{T}_{A\rightarrow A_{1}}$ be the
measurement map defined in equation~(\ref{measurementmap}). Let
$\mathcal{P}$ be the affine pair-wise independent permutation family defined
above. If $\{U_{1},\ldots,U_{|A|+1}\}$ defines a complete set of mutually
unbiased bases, then
\begin{align}
\frac{1}{|\mathcal{P}| (|A|+1)}\sum_{P_{\sigma}\in\mathcal{P}}\sum_{\theta=1}^{|A|+1} \Big|\Big| \hat{\Pi}_{A\rightarrow X_{1}}^{\theta\sigma}(\rho_{AE})    - \frac{\mathbb{I}_{A_{1}}}{|A_{1}|} \otimes \rho_{E} \Big|\Big|_1 \leq \varepsilon
\end{align}
where $\varepsilon = \sqrt{\frac{|A_{1}|}{|A|+1} 2^{-H_{\min}(A|E)}}$, $\hat{\Pi}_{A\rightarrow X_{1}}^{\theta\sigma}(\rho_{AE}) =\mathcal{T}_{A\rightarrow A_{1}}  \Big(P_{\sigma}U_{\theta}\rho_{AE}(P_{\sigma}U_{\theta})^{\dagger} \Big)$, and $\mathcal{P}=\{P_{\sigma}\}$ is a set of pair-wise independent permutation matrices. 

In particular, the set of unitaries $\{P_{\sigma}U_{\theta}: P_{\sigma}\in \mathcal{P},\theta\in [|A|+1] \}$ constitutes a $(k,\varepsilon)$-strong-QC-extractor provided:
\begin{equation}
\log_2 |A_{1}|\leq q+k-2\log_2(1/\varepsilon)
\end{equation}
with
\begin{equation}
L
=
(|A|+1)|\mathcal{P}|
=
(|A|+1)|A|(|A|-1)
\end{equation}
extraction unitaries.
\end{theorem}

\noindent\textit{Relation to the MUB-QCT security proof.}
Theorem~\ref{QCMUB} is the zero-smoothing specialization of
Theorem III.8 of Berta, Fawzi and Wehner \cite{QCE}. Its extraction
unitary is
$
P_\sigma U_\theta.
$
The  MUB-QCT preparation uses the adjoint unitary
$
V_{\theta\sigma}
=
(P_\sigma U_\theta)^\dagger.
$
Consequently, the preparation--measurement identity
(\ref{prepare-measure-duality}) maps every normalized adversarial POVM
effect directly to the QC-extractor source picture. This is the reason for
the adjoint convention adopted in Section~\ref{sec:MUBQCTProtocol}.

Theorem~\ref{QCMUB} therefore provides the general immediate-measurement
bound of Proposition~\ref{prop:QC-immediate}. The sharper binary bound of
Theorem~\ref{thm:immediate-measurement} additionally exploits the
rank-$d/2$ binary partition and the exact second-moment identity of the
complete MUB $2$-design.

The combination of a full set of MUBs and pair-wise independent permutations forms a \textit{strong} QC-randomness extractor. This strict property guarantees that even if the classical variables representing the choice of unitary $(U_{\theta}, P_{\sigma})$ are explicitly revealed to the adversary Eve \textit{after} the measurement $\hat{\Pi}_{A\rightarrow X_{1}}^{\theta\sigma}(\rho_{AE})$ is performed, the extracted bit remains statistically decoupled from Eve's composite system. Formally:
\begin{equation}
\Big|\Big| \rho_{X_{1}E U_{\theta}P_{\sigma}} - \frac{ \mathbb{I}_{A_{1}} }{|A_{1}|} \otimes \rho_{EU_{\theta}P_{\sigma}}  \Big|\Big|_1 \leq \sqrt{\frac{|A_{1}|}{|A|+1} 2^{-H_{\min}(A|E)}}
\end{equation}
where the fully classical-quantum joint state containing the revealed seed is defined as:
\begin{equation}
\rho_{X_{1}E U_{\theta}P_{\sigma}} := \Big( \frac{1}{(|A|+1)|\mathcal{P}|} \sum_{\theta=1}^{|A|+1} \sum_{P_\sigma \in \mathcal{P}} \hat{\Pi}_{A\rightarrow X_{1}}^{\theta\sigma}(\rho_{AE}) \otimes |\theta\rangle\langle \theta| \otimes |\sigma\rangle\langle  \sigma| \Big)
\end{equation}

\section{Single-Copy Error Rate Calculation}
\label{CSKR}

To evaluate the asymptotic secret key rate, we model the expected quantum bit error rate (QBER), denoted $p_e$, for a standard lossy fibre channel. Consider a lossy quantum channel with transmittance $T = 10^{-\alpha L/10}$, where $\alpha = 0.2 \text{ dB/km}$ is the attenuation coefficient and $L$ is the transmission distance. 

Let $\eta$ be the overall detector efficiency, $V$ the visibility of the detection setup (accounting for optical misalignment and intrinsic hardware errors), and $p_{\mathrm{dark}}$ the dark-count probability per detector per measurement window. We assume the standard low-photon and low-dark-count regime ($T\eta \ll 1$ and $p_{\mathrm{dark}} \ll 1$), allowing us to use first-order linear approximations for the detection probabilities.

Bob's measurement apparatus consists of two single-photon detectors. For a given channel use, let $D_c$ denote the ``correct'' detector corresponding to Alice's encoded bit $x$, and $D_w$ denote the ``wrong'' (orthogonal) detector.

\begin{enumerate}
\item \textbf{Signal detections:} The total probability of a detector clicking due to a successfully transmitted signal photon is $P_{signal} \approx T\eta$. Accounting for the visibility $V$, the probability that the signal correctly triggers $D_c$ is:
\begin{equation}
P(D_c\cap\text{signal}) \approx T\eta V
\end{equation}
Conversely, the probability that the signal erroneously triggers $D_w$ due to imperfect visibility is:
\begin{equation}
P(D_w\cap\text{signal}) \approx T\eta(1-V)
\end{equation}

\item \textbf{Dark counts:} The probability of a dark count occurring in a specific detector is $p_{\mathrm{dark}}$. Therefore, the probability of a noise-induced click in $D_c$ is $p_{\mathrm{dark}}$, and similarly for $D_w$. The total probability of a dark count occurring in either of the two detectors is:
\begin{equation}
P(\text{dark}) \approx 2 p_{\mathrm{dark}}
\end{equation}

\item \textbf{Total click probability:} The overall probability that Bob records a valid measurement event (a click in either detector) is the sum of the signal and dark count probabilities:
\begin{align}
P_{\mathrm{click}} \approx P_{signal} + P(\text{dark}) = T\eta + 2p_{\mathrm{dark}}
\end{align}
\end{enumerate}

An error occurs if Bob records a click in the incorrect detector $D_w$, which can be caused either by a misaligned signal photon or a dark count. The total probability of an erroneous click is:
\begin{align}
P_{\mathrm{wrong}} = P(D_w\cap\text{signal}) + P(D_w\cap\text{dark}) = T\eta(1-V) + p_{\mathrm{dark}}
\end{align}

Finally, the QBER $p_{e}$ is defined as the conditional probability of a wrong detection given that a click occurred in the apparatus:
\begin{equation}
p_{e} = \frac{P_{\mathrm{wrong}}}{P_{\mathrm{click}}} = \frac{T\eta(1-V) + p_{\mathrm{dark}}}{T\eta + 2p_{\mathrm{dark}}}
\label{pe}
\end{equation}

\section{Multi-Copy Error Rate and Abort Probability}
\label{App:MultiCopy}

To model the multi-copy detection probabilities discussed in
Section~\ref{sec:analysis}, suppose that Alice sends $m$ signal photons
through a channel of transmittance $T$ to detectors of efficiency $\eta$.
Conditioned on a signal photon being detected, it is routed independently
to the correct detector $D_c$ with probability $V$ and to the wrong detector
$D_w$ with probability $1-V$. We take $V$ to be independent of $m$; thus
misalignment is modeled as an i.i.d.\ per-photon process rather than as a
common-mode rotation of the complete $m$-photon signal. The dark-count
probability $p_{\mathrm{dark}}$ is applied once per detector and per protocol round,
corresponding to one detection gate covering the complete $m$-photon
signal.

Let
\begin{equation}
a:=T\eta V,
\qquad
b:=T\eta(1-V).
\end{equation}
For each transmitted photon, $a$ is the probability of a signal click in
$D_c$, $b$ is the probability of a signal click in $D_w$, and
$1-a-b=1-T\eta$ is the probability that the photon generates no signal
click.

Including one independent dark-count opportunity per detector and per
round, the probability that neither detector clicks is
\begin{equation}
P_{00}
=
(1-p_{\mathrm{dark}})^2(1-T\eta)^m.
\label{P00}
\end{equation}
The probabilities that $D_c$ or $D_w$, respectively, do not click are
\begin{align}
P(\overline D_c)
&=
(1-p_{\mathrm{dark}})(1-a)^m,
\\
P(\overline D_w)
&=
(1-p_{\mathrm{dark}})(1-b)^m.
\end{align}

Hence the probability of a correct accepted event, in which only $D_c$
clicks, is
\begin{equation}
P_{\mathrm{correct}}
=
P(\overline D_w)-P_{00}
=
(1-p_{\mathrm{dark}})(1-b)^m
-
(1-p_{\mathrm{dark}})^2(1-T\eta)^m,
\label{Pcorrect-m}
\end{equation}
while the probability of an erroneous accepted event is
\begin{equation}
P_{\mathrm{wrong}}
=
P(\overline D_c)-P_{00}
=
(1-p_{\mathrm{dark}})(1-a)^m
-
(1-p_{\mathrm{dark}})^2(1-T\eta)^m.
\label{Pwrong-m}
\end{equation}

The double-click probability is
\begin{equation}
P_{11}
=
1
-
P(\overline D_c)
-
P(\overline D_w)
+
P_{00}.
\end{equation}
Therefore
\begin{align}
P(\mathrm{abort})_m
&=
P_{00}+P_{11},
\\
QBER_m
&=
\frac{P_{\mathrm{wrong}}}
{P_{\mathrm{correct}}+P_{\mathrm{wrong}}}.
\end{align}
\end{document}